\documentclass{article}
\usepackage{graphicx}
\usepackage{amsmath}
\usepackage{amsfonts}
\usepackage{amssymb}
\newtheorem{theorem}{Theorem}

\begin{document}

\title{Two-dimensional models as testing ground for principles and concepts of local
quantum physics}
\author{Bert Schroer\\CBPF, Rua Dr. Xavier Sigaud 150 \\22290-180 Rio de Janeiro, Brazil\\and Institut fuer Theoretische Physik der FU Berlin, Germany}
\date{April 2005}
\maketitle
\tableofcontents
\begin{abstract}
In the past two-dimensional models of QFT have served as theoretical
laboratories for testing new concepts under mathematically controllable
condition. In more recent times low-dimensional models (e.g. chiral models,
factorizing models) often have been treated by special recipes in a way which
sometimes led to a loss of unity of QFT. In the present work I try to
counteract this apartheid tendency by reviewing past results within the
setting of the general principles of QFT. To this I add two new ideas: (1) a
modular interpretation of the chiral model Diff(S)-covariance with a close
connection to the recently formulated local covariance principle for QFT in
curved spacetime and (2) a derivation of the chiral model temperature duality
from a suitable operator formulation of the angular Wick rotation (in analogy
to the Nelson-Symanzik duality in the Ostertwalder-Schrader setting) for
rational chiral theories. The SL(2,Z) modular Verlinde relation is a special
case of this thermal duality and (within the family of rational models) the
matrix S appearing in the thermal duality relation becomes identified with the
statistics character matrix S. The relevant angular Euclideanization'' is done
in the setting of the Tomita-Takesaki modular formalism of operator algebras.

I find it appropriate to dedicate this work to the memory of J. A. Swieca with
whom I shared the interest in two-dimensional models as a testing ground for
QFT for more than one decade.

This is a significantly extended version of an ``Encyclopedia of Mathematical
Physics'' contribution hep-th/0502125. \ 
\end{abstract}

\section{Early history of two-dimensional solved models and the problem of
learning the right lessons}

Local quantum physics of systems with infinitely many interacting degrees of
freedom leads to situations whose understanding often requires new physical
intuition and mathematical concepts beyond that acquired in quantum mechanics
and perturbative constructions in quantum field theory. In this situation
two-dimensional soluble models turned out to play an important role. On the
one hand they illustrate new concepts and sometimes remove misconceptions in
an area where new physical intuition is still in process of being formed. On
the other hand rigorously soluble models confirm that the underlying physical
postulates are mathematically consistent, a task which for interacting systems
with infinite degrees of freedom is mostly beyond the capability of pedestrian
methods or brute force application of hard analysis on models whose natural
invariances has been mutilated by a cut-off.

In order to underline these points and motivate the interest in
two-dimensional QFT, let us briefly look at the history, in particular at the
physical significance of the three oldest two-dimensional models of relevance
for statistical mechanics and relativistic particle physics which are in
chronological order: the Lenz-Ising model, Jordan's model of
bosonization/fermionization and the Schwinger model (QED$_{2}$).

The Lenz-Ising (L-I) model was proposed in 1920 by Wihelm Lenz \cite{Lenz} as
the simplest discrete statistical mechanics model with a chance to go beyond
the P. Weiss phenomenological Ansatz involving long range forces and instead
explain ferromagnetism in terms of non-magnetic short range interactions. Its
one-dimensional version was solved 4 years later by his student Ernst Ising.
In his 1925 university of Hamburg thesis, Ising \cite{Ising} not only showed
that his chain solution could not account for ferromagnetism, but he also
proposed some (as it turned out much later) not entirely correct intuitive
arguments to the extend that this situation prevails to the higher dimensional
lattice version. His advisor Lenz as well as Pauli (at that time Lenz's
assistant) accepted these reasonings and as a result there was considerable
disappointment among the three which resulted in Ising's decision (despite
Lenz's high praise of Ising's thesis) to look for a career outside of research.

As a contribution to the many historical reminiscences on the occasion of the
2005 Einstein year, there is one episode which indirectly connects Einstein to
the beginnings of theoretical physics in Hamburg \footnote{I learned about
this episode through an email correspondence with K. Reich \cite{Reich}.}. The
university of Hamburg was founded in 1919, but the town fathers deemed it
unnessary to create a separate chair for theoretical physics. The newly
appointed mathematicians (Artin, Blaschke, Hecke) had the splendid idea to
invite Einstein (who at that time already enjoyed general fame) for a public
talk. The event was an overwhelming success, the talk on the concepts
underlying general relativity ended with a round of lively discussions,
especially questions coming from Hecke. The subsequent public pressure on the
city government led to a change of their decision. Einstein's indirect role as
a catalyzer was rapidly vindicated since the theoretical physics activities at
Hamburg university shortly thereafter led to the important discovery of the
exclusion principle and the construction of one of the most fruitful
two-dimensional models.

For many years a reference by Heisenberg \cite{Hei} (to promote his own
proposal as an improved description of ferromagnetism) to Ising's negative
result was the only citation; the situation begun to change when Peierls
\cite{Pei} drew attention to ``Ising's solution'' and the results of Kramers
and Wannier \cite{Kr-Wa} cast doubts on Ising's intuitive arguments beyond the
chain solution. The rest of this fascinating episode i.e. Lars Onsager's
rigorous two-dimensional solution exhibiting ferromagnetic phase transition,
Brucia Kaufman's simplification which led to conceptual and mathematical
enrichments (as well as later contributions by many other illustrious
personalities) hopefully remains a well-known part of mathematical physics
history even beyond my own generation.

This work marks the beginning of applying rigorous mathematical physics
methods to solvable two-dimensional models as the ultimate control of
intuitive arguments in statistical mechanics and quantum field theory. The L-I
model continued to play an important role in the shaping of ideas about
universality classes of critical behavior; in the hands of Leo Kadanoff it
became the key for the development of the concepts about order/disorder
variables (The microscopic version of the famous Kramers-Wannier duality) and
also of the operator product expansions which he proposed as a concrete
counterpart to the more general field theoretic setting of Ken Wilson. Its
massless version (and the related so-called Coulomb gas representation) became
a role model in the Belavin-Polyakov-Zamolodchikov \cite{B-P-Z} setting of
minimal chiral models and it remained up to date the only model with
non-abelian braid group commutation relations for which the n-point
correlators can and have been written down explicitly in terms of elementary
functions. Chiral theories confirmed the pivotal role of ``exotic'' statistics
in low dimensional QFT by exposing the appearance of \textit{plektonic}
statistics subject to the laws of Artin's braid group as a novel manifestation
of Einstein causality. As free field theories in higher dimensions are fixed
by their mass, spin and internal symmetries, the structure of chiral theories
is almost entirely determined by their braid group statistics data i.e. they
are in some sense as free (of genuine interactions, or as ``kinematical'') as
possible under the condition that they must realize nontrivial braid group
statistics. The latter is incompatible with a linear equation of motion and
the ensuing absence of correlations beyond two-point functions, a fact which
also has been established in the physically important case of braid group
statistics in d=1+2 where the statistics carrying fields are necessarily
semiinfinite string-like \cite{Mund1}.

Another conceptually rich model which lay dormant for almost two decades as
the result of a misleading speculative higher dimensional generalization by
its protagonist is the bosonization/fermionization model first proposed by
Pascual Jordan \cite{Jor1}. This model establishes a certain equivalence
between massless two-dimensional Fermions and Bosons; it is related to
Thirring's massless 4-fermion coupling model\footnote{In its original massless
and conformally invariant version it reached its mathematical perfection in
the work of Klaiber's Boulder lecture \cite{Klai}. Klaiber also enriched the
model by an additional parameter which allows a realization of the Thirring
model in a anyonic statistics mode with the expected spin-statistics relation
between the anomalous Lorentz spin and the anyonic commutation relation.} and
also to Luttinger's one-dimensional model of an electron gas \cite{Th}%
\cite{Lut}. One reason why even nowadays hardly anybody knows Jordan's
contribution (besides lack of comprehension of its content) is certainly the
ambitious but unfortunate title ``The neutrino theory of light'' under which
he published a series of papers; besides some not entirely justified criticism
of content, the reaction of his contemporaries consisted in a good-humored
carnivalesque Spottlied (mockery song) about its title \cite{Pais}. Its
mathematical content, namely the realization that the current of a free zero
mass Dirac fermion in d=1+1 is linear in canonical Boson creation/annihilation
operators (``Bosonization'') and that such zero mass Fermions (Jordan's
``neutrinos'') permit a formal representation in terms of ordered exponential
expressions in a free Boson field (``Fermionization'') turned out to play an
important illustrative role in the context of two-dimensional conformal
theories and their chiral decomposition. The massive version of the Thirring
model became the role model of integrable relativistic QFT and shed additional
light on two-dimensional Bosonization \cite{Co1}.

Both discoveries demonstrate the usefulness of having controllable
low-dimensional models; at the same time their complicated history also
illustrates the danger of rushing to premature ``intuitive'' conclusions about
extensions to higher dimensions. The search for the appropriate higher
dimensional analogon of a two-dimensional observation is an extremely subtle
but very worthwhile endeavour because if done correctly it often leads to
considerable conceptual progress. In the aforementioned two historical
examples the true physical message of those models only became clear through
hard mathematical work and profound conceptual analysis by other authors many
years after the discovery of the original model.

A review of the early conceptual progress through the study of solvable
two-dimensional models would be incomplete without mentioning Schwinger's
proposed solution of two-dimensional quantum electrodynamics, afterwards
referred to as the Schwinger model. Schwinger used this model in order to
argue that gauge theories are not necessarily tied to zero mass vector
particles; in opposition to the majority in the physics community (including
Pauli\footnote{For a long time it was thought that the use of abelian or
nonabelian gauge theories (as proposed by Yang and Mills and before (in 1938)
by Oscar Klein) in particle physics of massive vectormesons was not possible.
For this reason Sakurai presented his ideas about vector mesons without using
a gauge theoretical setting.}) he thought that it is conceivable that there
may exist a strong coupling regime of a QED-like gauge coupling which converts
the massless photons into a massive vectormeson and he used massless QED$_{2}$
to illustrate his point that the gauge theory setting does not exclude massive
vectormesons. His solution in terms of indefinite metric correlation functions
\cite{Schwinger} was however quite far removed from the interpretation of its
physical aspects. Some mathematical physics work and conceptual clarification
was necessary \cite{Lo-Sw} to unravel its physical content with the result
that the would-be charge of that QED$_{2}$ model was ``charged-screened'' and
hence its apparent chiral symmetry ``broken'', in other words the model exists
only in the so-called Schwinger-Higgs phase with massive free scalar particles
accounting for its physical content\footnote{As an illustration of historical
prejudices against Schwinger's ideas it is interesting to note that Swieca
apparently was not able to convince Peierls (on a visit to Brazil) about the
possibility of having massive vectormesons in a gauge theory (private
communication by J. A. Swieca around 1975).}). Another closely related aspect
of this model which also arose in the Lagrangian setting of 4-dimensional
gauge theories was that of the $\theta$-angle. Extended multi-component
versions of this model were used in for the study of problems of
charge-screening versus confinement \cite{A-R}. Although one believes that the
basic features of this difference between the Schwinger-Higgs screening
mechanism versus (fractional) fundamental flavor confinement continue to apply
in the 4-dimensional standard model, the lack of an intrinsic meaning of
notions of spin as well as statistics in massive d=1+1 models prevent
simple-minded analogies.

Thanks to its property of being superrenormalizable, the Schwinger model also
served as a useful testing ground for the Euclidean integral formulation in
the presence of Atiyah-Singer zero modes and their role in the Schwinger-Higgs
chiral symmetry breaking \cite{A-R}. These classical topological aspects of
the functional integral formulation attracted a lot of attention beginning in
the late 70s and through the 1980s but, as most geometrical aspects of the
Euclidean functional integral, their intrinsic physical significance remained
controversial. Even in those superrenormalizable 2-dim models, where the
measure theory underlying Euclidean functional integration can be
mathematically controlled \cite{Gl-Ja}, there is no good reason why within
such a setting topological properties derived from continuity requirements
should assert themselves outside of quasiclassical approximations
\cite{Jackiw}. This is no problem in the operator algebra approach where no
topological or differential geometrical property is imposed but certain
geometric structures (spacetime- and internal- symmetry properties) are
encoded in the causality and spectral principles of observable algebras.

In passing it is worthwhile to mention that Schwinger's idea on charge
screening found a rigorous formulation in a structural theorem which links the
issue of charge (carried by physical particles) versus charge screening to the
spectral property near zero of the mass operator \cite{Sw}. Mass generation
via charge screening in 4-dimensional \textit{perturbation} theory is not
possible without additional physical (Higgs) degrees of freedom.

\ The most coherent and systematic attempt at a mathematical control of
two-dimensional models came in the wake of Wightman%
\'{}%
s first rigorous programmatic formulation of QFT \cite{St-Wi}. This
formulation stayed close to the ideas underlying the impressive success of
renormalized QED perturbation theory, although it avoided the direct use of
ideas of Lagrangian quantization. The early attempts towards a ``constructive
QFT'' found their successful realization in two-dimensional QFT (the
$P\varphi_{2}$ models \cite{Gl-Ja}). Only in low dimensional theories the
presence of Hilbert space positivity and energy positivity can be reconciled
with the kind of mild short distance singularity behavior
(super-renormalizability) which the methods of constructive QFT requires.
Despite interesting later additions after the appearance of the cited book,
this barrier has essentially persisted. For this reason the main attention
will be focussed on alternative constructive methods which are free of this
restrictions; they have the additional advantage to reveal more details about
the conceptual structure of QFT beyond the assertion of their existence. The
best illustration of the constructive power of these new methods comes from
massless d=1+1 conformal and chiral QFT as well as from massive factorizing
models. Their presentation and that of the contained messages for general QFT
will form the backbone of this article.

There are several books and review articles \cite{F-S-T}\cite{Gi}\cite{D-M-S}
on d=1+1 conformal as well as on massive factorizing models \cite{A-R}%
\cite{Olalla}. To the extend that concepts and mathematical structures are
used which permit no known generalization to higher dimensions (e.g. Kac-Moody
algebras, loop groups, integrability, presence of an infinite number of
conservation laws), their line of approach will not be followed in this report
since our primary interest will be the use of two-dimensional models of QFT as
``theoretical laboratories'' of general QFT as stated in the abstract.

Our aim is two-fold; on the one hand we intend to illustrate known principles
of general QFT in a mathematically controllable context and on the other hand
we want to identify new concepts whose adaptation to QFT in d=1+1 lead to
their solvability. In emphasizing the historical side of the problem by using
the oldest solved two-dimensional models as a vehicle for the introduction of
relevant concepts, I also hope to uphold the awareness of the unity and
historical continuity in QFT in times of rapidly changing fashions in the age
of electronic communications.

\section{General concepts and their two-dimensional adaptation}

The general framework of QFT, to which the rich world of controllable
two-dimensional models contributes as a testing ground, exists in two quite
different but nevertheless closely related formulations: the setting in terms
of pointlike covariant fields due to Wightman \cite{St-Wi}, and the more
algebraic setting initiated by Haag and Kastler based on spacetime-indexed
operator algebras \cite{Haag} and concepts which developed over a long period
of time with contributions of many other authors. Whereas the Wightman
approach aims directly at the (not necessarily observable) quantum fields, the
operator algebraic setting is more ambitious. It starts from physically
motivated assumptions about the algebraic structure of local (spacelike
commuting) observables\footnote{The minimal requirement on observable fields
or localized algebras is that they are ``bosonic'' i.e. commute for spacelike
distances whereas a maximal definition would require the absence of any
internal symmetry.} and structurally reconstructs (i.e. demonstrates
rigorously that all concepts and implementing objects of particle physics are
in place) the full field theory (including the operators carrying the
superselected charges) in the spirit of a local representation theory of the
assumed structure of the local observables. This has the advantage that the
somewhat mysterious concept of an inner symmetry (as opposed to ``outer''
(spacetime) symmetry) can be traced back to its physical roots which is the
representation theoretical structure of the local observable algebra. In the
Lagrangian quantization approach the inner symmetry is part of the input (the
multiplicity index of field components on which subgroups of SU(n) or SO(n)
act linearly) and hence it is not possible to even formulate this fundamental
question. Whenever the sharp separation (the Coleman-Mandula theorems
\cite{Co-Ma}) of inner versus outer symmetry becomes blurred as a result of
the appearance of braid group statistics in low spacetime dimensions, the
Lagrangian quantization setting is inappropriate and even the Wightman
framework has to be extended. In that case the algebraic approach is the most appropriate.

The two most important physical properties which are shared between the
Wightman approach (WA) \cite{St-Wi} and the operator algebra (AQFT) setting
\cite{Haag}, are the spacelike locality (often referred to as Einstein
causality) and the restriction to the stability ensuring positive energy
representations of the Poincar\'{e} group which implement the covariance of
the Wightman fields respectively the local observable algebras.

\begin{itemize}
\item  Spacelike commutativity of quantum fields or of local observable
algebras:
\begin{align}
&  \left[  \psi(x),\varphi(y)\right]  _{\mp}=0,\;(WA)\\
&  \mathcal{A}(\mathcal{O}^{\prime})\subseteq\mathcal{A}(\mathcal{O})^{\prime
},\,\mathcal{O}\text{ }open\,\,nbhd.\;(AQFT)\nonumber
\end{align}

\item  Positive energy reps. of the Poincar\'{e} group $\mathcal{P}:$%
\begin{align}
&  U(a,\Lambda)\psi(x)U(a,\Lambda)^{\ast}=D^{-1}(\Lambda)\psi(\Lambda
x+a),\,\,(WA)\\
&  U(a,\Lambda)\mathcal{A}(\mathcal{O})U(a,\Lambda)^{\ast}=\mathcal{A}%
(\mathcal{O}_{(a,\Lambda)}),\,\,(AQFT)\nonumber\\
&  U(a)=e^{iPa},\text{ }specP\subseteq V_{+},\,\,P\Omega=0\nonumber
\end{align}
\end{itemize}

Here $\psi(x),\varphi(x)...$are (singular) field operators (operator-valued
distributions) in a Wightman QFT which are assumed to either commute or
anticommute for spacelike distances and a structural theorem \cite{St-Wi} ties
the commutator relation to finite dimensional representations of the Lorentz
group $\mathcal{L}$ whereas the anticommutator has to be used for projective
representations (which turn out to be usual vector-representations of the
two-fold covering $\widetilde{\mathcal{L}}).$ The observable algebra consists
of a family of (weakly closed) operator algebras $\left\{  \mathcal{A(O)}%
\right\}  _{\mathcal{O\in K}}$ indexed by a family of convex causally closed
spacetime regions $\mathcal{O}$ (with $\mathcal{O}^{\prime}$ denoting the
spacelike complement and $\mathcal{A}^{\prime}$ the von Neumann commutant)
which act in one common Hilbert space; it is suffient to know these local
algebras in the vacuum representation i.e. without loss of generality one can
identify $\mathcal{A(O)}$ with $\mathcal{\pi}_{0}(\mathcal{A(O)})$. Certain
properties cannot be naturally formulated in the pointlike field setting (e.g.
Haag duality for simply connected causally complete regions $\mathcal{A}%
(\mathcal{O}^{\prime})=\mathcal{A}(\mathcal{O})^{\prime}),$ but the connection
between the two formulations of local quantum physics is nevertheless quite
close; in particular in case of two-dimensional theories there are convincing
arguments that one can pass between the two without having to impose
additional technical requirements. There exists a recent generalization of
this algebraic framework which incorporates the Einstein local covariance
principle in which the above setting re-emerges as a special case
\cite{B-F-V}. In section 5 we will present a chiral illustration of these new concepts.

The above two requirements are often (depending on the kind of structural
properties one wants to derive) complemented by additional impositions which,
although not carrying the universal weight of principles, nevertheless
represent natural assumptions whose violation (even though not prohibited by
the principles) would cause paradigmatic attention and warrants special
explanations. Examples are ``weak additivity'', ``Haag duality'' and ``the
split property''. Weak additivity i.e. the requirement $\vee\mathcal{A(O}%
_{i}\mathcal{)}=\mathcal{A(O)}$ if $\mathcal{O}=\cup\mathcal{O}_{i}$ expresses
``the global results from amalgamating the local'' aspect which is inherent in
the pointlike field formulation, but needs to be added in the algebraic approach.

Haag duality is the statement that the commutant not only contains the algebra
of the causal complement (Einstein causality), but is even equal to it, i.e.
$\mathcal{A}(\mathcal{O}^{\prime})=\mathcal{A}(\mathcal{O})^{\prime}.$ Its
obvious connection with the measurement process (it assigns a localization to
measurements which are commensurable with the totality of all measurements
which are performable in a prescribed spacetime region $\mathcal{O}$) suggests
to look for a profound physical explanation whenever it is violated e.g. its
violation for observables localized in a simply connected causally complete
region signals spontaneous symmetry breaking in the associated charge-carrying
field algebra \cite{Haag}. The possibilities of spontaneous symmetry breaking
in d=1+1 are severely restricted. The Bisognano-Wichmann property for wedge
localized algebras (\ref{mod}) assures that Haag duality can be enforced by a
symmetry-reducing (to the unbroken subgroup) extension via dualization. In
conformal theories it is always satisfied independent of spacetime dimension
\cite{B-G-L1}. Its violation for multi-local region reveals the charge content
of the model by enforcing charge-anticharge splittings in the neutral
observable algebra \cite{K-L-M}.

The split property \cite{Do-Lo}, namely the algebraic isomorphism to a quantum
mechanical type tensor factorization $\mathcal{A}(\mathcal{O}_{1}%
\cup\mathcal{O}_{2})\simeq\mathcal{A}(\mathcal{O}_{1})\otimes\mathcal{A}%
(\mathcal{O}_{2})$ for regions $\mathcal{O}_{i}$ with a (arbitrarily small)
spacelike separation between them, is a result of the adaptation of the
``finite number of degrees of freedom per unit cell in phase space''
requirement of QM to QFT which leads to the so-called ``nuclearity property''
\cite{Haag}. Looming behind all these properties is the inexorable vacuum
polarization; in order to prevent its infinity-creating reponse to sharp
localization which destroys the strong property of quantum mechanical
tensor-factorization for nonoverlapping regions, one needs the finite
spacelike distance required by the split property.

Related to the Haag duality is the local version of the ``time slice
property'' (the QFT counterpart of the classical causal dependency property)
sometimes referred to as ``strong Einstein causality'' $\mathcal{A}%
(\mathcal{O}^{\prime\prime})=\mathcal{A}(\mathcal{O})^{\prime\prime}.$

One of the most astonishing achievements of the algebraic approach is the DHR
theory of superselection sectors \cite{DHR} i.e. the realization that the
structure of charged (nonvacuum) representations (with the unrestricted
superposition principle being valid only within one representation) and the
spacetime properties of the fields which are the carriers of these generalized
charges including their spacelike commutation relation (leading to the
particle statistics and to internal symmetry \cite{D-R}) are already encoded
in the structure of the Einstein causal observable algebra. The intuitive
basis of this remarkable result (whose prerequisite is causal locality) is
that one can generate charged sectors by spatially separating charges in the
vacuum (neutral) sector and disposing of the unwanted charges at spatial
infinity \cite{Haag}. In higher spacetime dimensions the charge fusion
structure turns out to be isomorphic to the tensor product operation of
compact group representations; the framework does not exclude any compact group.

An important concept which especially in d=1+1 has considerable constructive
clout is ``modular localization''. It is a consequence of the above algebraic
setting if either the net of algebras has pointlike field generators, or if
the one-particle masses are separated by spectral gaps so that the formalism
of time dependent scattering can be applied \cite{Mund2}; in conformal
theories this property holds automatically (in all spacetime dimensions). It
rests on the basic observation \cite{Summers} that a \textit{standard pair}
($\mathcal{A},\Omega$) of a von Neumann operator algebra and a
vector\footnote{Standardness means that the operator algebra of the pair
($\mathcal{A},\Omega$) act cyclic and separating on the vector $\Omega.$}
gives rise to a Tomita operator S through its star-operation whose polar
decomposition yield two modular objects, a 1-parametric subgroup of the
unitary group of operators in Hilbert space $\Delta^{it}$ whose Ad-action
defines the modular automorphism of ($\mathcal{A},\Omega$) whereas the angular
part $J$ is the modular conjugation which maps $\mathcal{A}$ into its
commutant $\mathcal{A}^{\prime}$
\begin{align}
&  SA\Omega=A^{\ast}\Omega,\;\;S=J\Delta^{\frac{1}{2}}\label{mod}\\
J_{W}  &  =U(j_{W})=S_{scat}J_{0},\;\Delta_{W}^{it}=U(\Lambda_{W}(2\pi
t))\nonumber\\
&  \sigma_{W}(t):=Ad\Delta_{W}^{it}\nonumber
\end{align}
The standardness assumption is always satisfied for any field theoretic pair
($A(\mathcal{O}),\Omega$) of a $O$-localized algebra and the vacuum state (as
long as $\mathcal{O}$ has a nontrivial causal disjoint $\mathcal{O}^{\prime}$)
but it is only for the wedge region $W$ that the modular objects have a
physical interpretation in terms of the global spacetime symmetry group of the
vacuum as specified in the second line (\ref{mod}), namely the modular unitary
represents the $W$-associated boost $\Lambda_{W}(\chi)$ and the modular
conjugation implements the TCP-like reflection along the edge of the wedge
\cite{Bi-Wi}. The third line is the definition of the modular group. Its
usefulness results from the fact that it does not depend on the state vector
$\Omega$ but only on the state $\omega(\cdot)=\left(  \Omega,\cdot
\Omega\right)  $ which it induces. Another noteworthy fact is that the modular
group $\sigma^{(\eta)}(t)$ associated with a different state $\eta(.)$ is
unitarily equivalent to $\sigma^{(\omega)}(t)$ with a unitary $u(t)$ which
fulfills the Connes cocycle property. The importance of this theory for local
quantum physics results from the fact that it leads to the concept of
\textit{modular localization\footnote{As opposed to classical localization via
spacetime support properties of functions. Often one can construct intertwiner
which transform quantum (modular) localization into classical localization
(e.g. the u-v interwiners which lead from Wigner creation/annihilation
operators to local point- or string-like covariant ``fields''), but in general
the relation between nets of spacetime localized algebras and their
operator-distribution valued generators is quite subtle. }}, a new totally
intrinsic (i.e. independent of field coordinatizations) scenario for field
theoretic constructions which is different from the Lagrangian quantization schemes.

A good starting point for understanding the physical aspects and aims of
modular localization is the Wigner particle representation theory.
Localization in analogy to the Born probability interpretation in quantum
mechanics is incompatible with relativistic covariance since there are simply
no covariant localizing projection operators (even if one extends the Wigner
space to the Fock space). A recent review of these facts and the physics
behind the concept of modular localization which replaces the concept of
localizing via projection operators can be found in \cite{cross2}. The
construction of modular localized subspaces of a positive energy Wigner space
$H^{(1)}$ starts from the group theoretic definition of a wedge-affiliated
S-operator $S_{W}^{(1)}$ by multiplying the (unbounded) analytically continued
wedge affiliated boost $\Delta_{W}^{(1)}\equiv U_{W}^{(1)}(\Lambda_{W}\left(
-\pi i\right)  )$ with the antiunitary involution $J_{W}^{(1)}$ which
represents the reflection along the edge of the wedge as in (\ref{mod}) but
without an operator algebra being present. As a result of the commutation of
$\Lambda_{W}$ and $j_{W}$ and the antilinearity of $J_{W}^{(1)}$ this
unbounded involutive antiunitary operator in $H_{{}}^{(1)}$ fulfills all the
properties of a Tomita S-operator and its +1 eigenspace defines a real
subspace $K_{W}^{(1)}$ of the complex Wigner space. The sharpening of
localization is achieved by intersecting $K_{W}^{(1)}{}^{\prime}s$ or, what
amounts to the same, via directly defining new S-operators by intersecting
domains of definition of $S_{W}s$\cite{Mund3}. For finite spin representations
the intersections associated with (compact) double cone regions $K_{D}%
^{(1)}=\cap_{W\supset D}K_{W}^{(1)}$ are nontrivial and ``standard'' (implying
that $K_{D}^{(1)}+iK_{D}^{(1)}$ is dense in $H^{(1)}$), but for zero mass
infinite spin representations as well as for massive d=1+2 anyonic spin
representations these intersections are trivial and the smallest nontrivial
and standard intersections are (noncompact) spacelike cones with semiinfinite
stringlike cores. The important theorems on modular localization within a
group representation setting can be found in \cite{B-G-L2}.

The transition from modular localized subspaces to localized operator algebras
of interaction-free systems is done in a functorial way using the Weyl (or CAR
in case of halfinteger spin) operators which map the modular localized
function spaces into spacetime-indexed operator algebras (generated by the
images of the functor). The functorial relation between modular localized
subspaces and localized von Neumann subalgebras\footnote{The $\Delta_{W}%
^{it},J_{W}$ and $S_{W}$ operators in Fock space are related to their
one-particle analogs by the rules of second quantization.} commutes with more
stringent localization which is achieved by intersecting wedge spaces or wedge
algebras. Point- and string-like generators (necessarily singular i.e.
distribution-valued) can be constructed with the help of intertwiners (the
analog of $u$ and $v$ spinors) which relate the modular (quantum) localization
to the classical localization in terms of test function supports \cite{F-S}
\cite{Mund3} \cite{M-S-Y}. The concept of modular localization solves the
age-old problem of the continuous spin Wigner representation by showing that
the compact localization spaces in those cases (and also for d=1+2 anyons) are
empty; in those cases the tightest possible localization is in (arbitrarily
thin) spacelike cones associated with stringlike distributional generators.
The application of modular localization to the Wigner one-particle spaces
becomes especially simple for massless two-dimensional theories (since as a
result of their decomposability into two chiral components the energy-momentum
spectrum is not subject to a mass-shell restriction).

For interacting systems the construction of spacetime-indexed operator
algebras looses its functorial relation with modular localization of Wigner
particle states \cite{cross2}. Instead of being geometrically defined in terms
of Poincar\'{e} group representation theory, modular localized subspaces in
Fock space are simply identified with the dense subspaces $\mathcal{A(O)}%
\Omega$ which the local operator algebras generate from the vacuum after they
have been closed in the graph topology of the Tomita S-operator of the
standard pair ($\mathcal{A(O),\Omega}$) i.e. $H(\mathcal{O})\equiv
\overline{\mathcal{A(O)}\Omega}_{graph(S)}=\left\{  \psi|\,\left|
\psi\right|  ^{2}+\left|  S\psi\right|  ^{2}<\infty\right\}  ,$ $H(O)\overset
{dense}{\subset}H.$ As a result of geometric simplifications the application
of modular theory to two dimensional theories leads to particularly powerful results.

For the later purpose of analyzing ``thermal duality'' we define what is meant
by ``Euclideanization'' of the modular structure. We simply change the Hilbert
space inner product by defining the following positive definite sesquilinear
form \cite{Euclid} on the dense set of state vectors $L_{\mathcal{A}}%
\equiv\mathcal{A}\Omega$
\begin{equation}
\left\langle \Xi(A\Omega),\Xi(B\Omega)\right\rangle :=(A\Omega,\Delta
^{\frac{1}{2}}B\Omega)=(\Omega,JAJB\Omega) \label{Eu}%
\end{equation}
whose closure defines (thanks to the properties of the modular objects) the
new ``Euclidean'' Hilbert space $H^{E}$ (the map $\Xi$ in $L_{\mathcal{A}}%
^{E}\equiv\Xi(L_{\mathcal{A}})$ denotes the Euclidean re-interpretation) on
which this changed inner product leads to a new $^{\dagger}$Euclidean star
algebra by starting from the formula
\begin{align}
&  \left\|  \Xi(A)B\Omega\right\|  _{H^{E}}^{2}=\left\|  \Delta^{\frac{1}{4}%
}AB\Omega\right\|  _{H}^{2}\leq\left\|  \Delta^{\frac{1}{4}}A\Delta^{-\frac
{1}{4}}\right\|  _{H}^{2}\left\|  B\right\|  _{H^{E}}^{2}\\
&  D(\sigma_{-\frac{i}{4}})\equiv\left\{  A\in\mathcal{A\,}|\,\Delta^{\frac
{1}{4}}A\Delta^{-\frac{1}{4}}\in B(H)\right\}  ,\text{ }\nonumber\\
&  \curvearrowright\Xi(D(\sigma_{-\frac{i}{4}}))\subset B(H^{E})\nonumber
\end{align}
where the second line consists of the definition on a shared subalgebra
$A_{sh}=D(\sigma_{-\frac{i}{4}})$ i.e. an algebra of bounded operators
(without a star operation) which belongs to both the original setting and its
Euclidean companion. It contains the dense subalgebra $\mathcal{A}_{an}$ of
$\sigma_{t}$-analytic elements and affiliated pointlike field generators.
Equipped with the original $^{\ast}$star, the von Neumann double commutant in
$H$ is equal to the original operator algebra $A=(A_{sh})_{\ast}^{\prime
\prime},$ while using instead the Euclidean $\dagger$star the double commutant
in $H^{E}$ defines the Euclidean algebra $\mathcal{A}^{E}=\left(
A_{sh}\right)  _{\dagger}^{\prime\prime}$.

It is easy to see\ that ($\mathcal{A}^{E},\Omega^{E}$) defines again a
standard pair with $\Delta_{E}^{it}=\Delta^{it},$ whereas the old star becomes
the new $J^{E}$ action and the old $J$-action the new star. In the physical
$\mathcal{A}(W)$ situation one would phrase this interchange by saying that
the $\ast$-conjugation (related to the physical charge-conjugation) and the
$J$-reflection are interchanged. The use of the terminology ``Euclidean''
becomes clear if one specializes this formalism to chiral theories on the
lightray; in that case the effect of the change of the inner product on the
one-sided translation $U(x)$ of the right wedge (rather halfline) algebra
$\mathcal{A}(0,\infty)$ leads to the formula
\begin{equation}
\Delta^{\frac{1}{4}}U(x)\Delta^{-\frac{1}{4}}=U^{E}(x)=\text{ }``U(ix)"
\label{trans}%
\end{equation}
(which has a meaning with respect to $\mathcal{A}_{sh}(0,\infty),$ the
``starless'' algebra) which mediates between $\mathcal{A}(0,\infty)$ and
$\mathcal{A}^{E}(0,\infty).$ More details will be deferred to a separate paper.

A variant of this modular analog of the Osterwalder-Schrader property, which
uses instead of a one- a two-sided compression $U(x)$ for $x>0$ a suitably
defined 2-sided compression on $\mathcal{A}(0,\infty),$ is the crucial
structure behind the thermal duality and the Verlinde relation in subsection 5.

A special property of d=1+1 Minkowski spacetime is the disconnectedness of the
right/left spacelike region which leads to a right-left ordering structure. So
in addition to the Lorentz invariant timelike ordering $x\prec y$ (x earlier
than y) which exists in any spacetime dimension, there is an invariant
spacelike ordering\footnote{The left/right ordering defines a class division
of pairs (x,y) under causality-preserving changes.} $x<y$ (x to the left of y)
in d=1+1; this opens the possibility of more general Lorentz-invariant
spacelike commutation relation than those implemented by Bose/Fermi fields
e.g. \textit{plektonic} fields with a spacelike Artin braid group commutation
structure. The appearance of such exotic statistics fields is not compatible
with their Fourier transforms being on-shell creation/annihilation operators
for Wigner particles, rather the states they generate from the vacuum contain
in addition to the one-particle contribution a vacuum polarization cloud
\cite{Mund1}. This close connection between new kinematic possibilities and
interactions is one of the reasons why, different from higher dimensions
(where interactions are prescribed by recipes based on local couplings of free
fields, usually within the setting of Lagrangian quantization) low dimensional
QFT offers an easier and more intrinsic access to the central issue of
interactions. Although the operator-algebraic formulation is particularly
well-suited to a more intrinsic approach, this does not mean that pointlike
covariant fields have become obsolete. They only changed their role; instead
of mediating between classical and quantum field theory in the (canonical or
functional integral) setting of Lagrangian quantization, they are now
universal generators of all local algebras and hence also of all modular
objects which taken together generate an infinite dimensional noncommutative
universal unitary group in the Hilbert space. Besides the implementors of
global spacetime symmetries this universal modular group also contains
\ (section 5) ``partial diffeomorphisms'' whose modular generated unitaries
only act geometrically on localized subalgebras.

\section{Boson/Fermion equivalence and superselection theory in a special model}

The simplest and oldest but yet conceptually quite rich model is obtained (as
first proposed by Pascual Jordan \cite{Jor1}) by using a 2-dim. massless Dirac
current and showing that it may be expressed in terms of scalar canonical Bose
creation/annihilation operators\footnote{The bilinear formulae which relate
these operators to the original Fermion operators can be found in
\cite{Klai}.}
\begin{equation}
j_{\mu}=:\overline{\psi}\gamma_{\mu}\psi:=\partial_{\mu}\phi,\,\phi
:=\int_{-\infty}^{+\infty}\{e^{ipx}a^{\ast}(p)+h.c.\}\frac{dp}{2\left|
p\right|  }%
\end{equation}
Although the potential of the current as a result of its infrared divergence
is not a field in the standard sense of an operator-valued distribution in the
Fock space of the a(p)$^{\ast}$ creation/annihilation operators \footnote{It
becomes an operator after smearing with test functions whose Fourier transform
vanishes at p=0.}$,$ the formal exponential defined as the zero mass limit of
a well-defined exponential free massive field (taken inside correlation
functions)
\begin{equation}
:e^{i\alpha\phi(x)}:=lim_{m\rightarrow0}m^{\frac{\alpha^{2}}{2}}%
:e^{i\alpha\phi_{m}(x)}: \label{mass}%
\end{equation}
turns out to be a well-defined quantum (i.e. infrared finite) quantum field.
Here the limit is understood in the sense of vacuum expectation values using
the Wick combinatorics of the massive free field; the power in the
pre-exponential mass factor is determined by the requirement that the most
singular contribution from the Wick contraction stays finite. For example the
leading singular part in $m\rightarrow0\,$of the two-point functions behaves
as
\begin{align}
&  \left\langle :e^{i\alpha\phi_{m}(x)}::e^{\mp i\alpha\phi_{m}(y)}%
:\right\rangle \sim m^{\mp\alpha^{2}}\left(  \frac{1}{\left(  x-y\right)
_{\varepsilon}^{2}}\right)  ^{\pm\alpha^{2}}\label{2-point}\\
&  \frac{1}{\left(  \xi\right)  _{\varepsilon}^{2}}:=lim_{\varepsilon
\rightarrow0}\frac{1}{\xi^{2}+i\varepsilon sign\xi_{0}}=lim_{\varepsilon
\rightarrow0}\frac{1}{\xi_{+}+i\varepsilon}\frac{1}{\xi_{-}+i\varepsilon
}\nonumber
\end{align}
where in the last line the leading power has been re-written in lightray
coordinates $\xi_{\pm}$ (with the correct positive energy $i\varepsilon$
-prescription). In order to maintain a finite zero mass limit one must use in
(\ref{mass}) precisely that mass power which keeps all all correlations
finite. Thanks to the general Wick combinatorics of exponential fields this
leads to ($\xi_{ij}\equiv x_{i}-x_{j}$)%
\begin{equation}
\left\langle \prod_{i}:e^{i\alpha_{i}\phi(x)}:\right\rangle =\left\{
\begin{array}
[c]{c}%
\prod_{i<j}\left(  \frac{-1}{\left(  \xi_{+ij}\right)  _{\varepsilon}\left(
\xi_{-ij}\right)  _{\varepsilon}}\right)  ^{\frac{1}{2}\alpha_{i}\alpha_{j}%
},\,\sum\alpha_{i}=0\\
0,\,\,\,\,\,\,otherwise
\end{array}
\right.  \label{cons}%
\end{equation}
i.e. those correlations which which do not obey the \textit{charge selection
rule} $\sum\alpha_{i}=0$ vanish and the nonvanishing ones factorize into
chiral components i.e. the model splits into two identical \ independent
chiral theories. The additional presence in the vacuum expectation values of
an arbitrary polynomial in the current $\prod_{i}j_{\mu}{}_{i}(y_{i})$ would
not change these arguments which insures that the resulting zero mass limiting
theory is a bona fide quantum field theory i.e. its system of Wightman
functions is canonically associated (via the GNS construction) with an
operator theory in a Hilbert space with a distinguished vacuum vector. There
exists another physically more intuitive and intrinsic method (whose
mathematical formulation is more involved) where one stays in the zero mass
setting and obtains the charged sectors by splitting neutral states (belonging
to the vacuum sector) and ``dumping the unwanted compensating charge behind
the moon'' \cite{Haag} (i.e. one uses spacial infinity as a wast-disposal). In
that case one starts from smeared exponential $expij(f)$ of the current with
smearing functions $f$ which are the smoothened version of a characteristic
function $\chi(x,a),$ so that formally they represent exponentials of
$\phi(x)$ smeared with bilocal function $\partial f$ being supported around
$x$ (with positive values) and $a$ (with negative values) such that the total
integral vanishes. The properly renormalized exponential
\begin{equation}
\frac{1}{Z(f)}e^{i\alpha j_{0}(f)}%
\end{equation}
has a finite limit for $a\rightarrow\infty$ in spacelike direction (within
vacuum expectation values) precisely if the charge conservation among all
finite endpoints $x_{i}$ in products of such smeared exponentials is
maintained. The mechanism resembles the previous argument; the contributions
from the contractions from the finite endpoints with the ends going to
infinity lead to a vanishing result in case of no charge compensation between
the finite ends (and approach a nontrivial finite limit with charge
compensation). A conceptually very attractive method which determines the
charge content without the disposal of the unwanted charge at infinity can be
formulated in terms of inclusions which are canonically associated to disjoint
two-intervals \cite{K-L-M} (see next subsection).

The abelian current model and its associated charge-carrying exponential
fields permit an extension to the compactified lightray $R\rightarrow\dot
{R}\equiv S^{1}$ which is done with the help of the Cayley transformation%

\begin{equation}
z=\frac{1+ix}{1-ix}\in S^{1} \label{Cayley}%
\end{equation}
i.e. the Wightman fields which are operator-valued Schwartz distributions can
be extended to a larger test function space which consists of smooth functions
on the circle without the infinite order zero at $z=-1$ which corresponds to
the fast decrase at $x=\pm\infty$). The structural property which places this
extendibility on more general model-independent footing is conformal
invariance; its more systematic exploration will be the subject of the two
next subsections where it will become clear that the chiral decomposition into
two lightray theories is a general property of two-dimensional conformal
covariant theories.

There is a fine point in the compactified description namely the occurrence of
a quantum mechanical zero modes in the Fourier decomposition of the circular
description. It is not difficult to verify that their presence leads to a
quantum mechanical pre-exponential factor for the Wick-ordered exponential
fields which automatically enforce the $\alpha$-charge conservation. Hence in
the rotational description the Wick contraction formalism holds in a standard
way without having to add charge conservation by hand as in (\ref{cons}); they
simply result from the zero mode quantum mechanics. In this approach the
original chiral Dirac Fermion $\psi(x)$ from which the (chiral component of
the current) was formed as a $j=\psi^{\ast}\psi$ composite re-appears as a
charge-carrying exponential field for $\alpha=1$ and illustrates the meaning
of bosonization/fermionization. But note that this terminology has to be taken
with a grain of salt in view of the fact that the bosonic current algebra only
generates a superselected subspace of the space generated by the iterative
application charge-carrying exponential fields i.e. although the Boson lives
in the Fermion sector, the Fermion operator creates state vectors which are
outside the Bose vacuum sector. Only in massive two-dimensional theories a
complete bosonization/fermionization (in which the generated spaces are
identical) can be achieved, a problem which is related to the irrelevance of
statistics and the appearance of order/disorder variables in such massive
models (see last subsection).

It is amusing to note that Jordan's treatment of fermionization had such a
pre-exponential quantum mechanical factor. At this point it should however be
clear to the reader that the physical content of Jordan's paper had nothing to
do with its misleading title ``neutrino theory of light'' but rather was a
special illustration about charge superselection rules in QFT long before this
general concept was recognized.

A systematic approach which avoids pointlike fields in favor of spacetime
indexed operator algebras can be formulated in terms of positive energy
representation theory for the Weyl algebra\footnote{The Weyl algebra was not
used in QFT at the time of Jordan's paper. By representation we mean here a
regular representation in which the exponentials can be differentiated in
order to obtain (unbounded) smeare current operators.} on the circle (which is
the rigorous operator algebraic formulation of the abelian current algebra).
It is the operator algebra generated by the exponential of a smeared chiral
current (always with real test functions) with the following relation between
the generators
\begin{align}
&  W(f)=e^{ij(f)},\;\,j(f)=\int\frac{dz}{2\pi i}j(z)f(z),\,\left[
j(z),j(z^{\prime})\right]  =-\delta^{\prime}(z-z^{\prime}),\;\\
&  W(f)W(g)=e^{-\frac{1}{2}s(f,g)}W(f+g),\,\,\,W^{\ast}(f)=W(-f)\nonumber\\
&  \mathcal{A}(S^{1})=alg\left\{  W(f),f\in C_{\infty}(S^{1})\right\}
,\mathcal{A}(I)=alg\left\{  W(f),suppf\subset I\right\} \nonumber
\end{align}
where $s(f,g)=\int\frac{dz}{2\pi i}f^{\prime}(z)g(z)$ is the symplectic form
which characterizes the Weyl algebra structure and the last line denotes the
unique $C^{\ast}$ algebra generated by the unitary objects $W(f).$ A
particular representation of this algebra is given by assigning the vacuum
state on the generators
\begin{align}
&  \left\langle W(f)\right\rangle _{0}=e^{-\frac{1}{2}\left\|  f\right\|
_{0}^{2}},\,\left\|  f\right\|  _{0}^{2}=\sum_{n\geq1}n\left|  f_{n}\right|
^{2}\label{vac}\\
&  L_{0}=\int\frac{dz}{2\pi i}T(z),\,\,T(z)=:j(z)^{2}:\nonumber
\end{align}
The norm in the first line leads to an inner product space which can be made
into a Hilbert space by defining a complex structure\footnote{The
multiplication with i which characterizes a complex struture in the present
case is multiplication with the number $\pm i$ of the $\pm$ Fourier
components.} on the real space. In the last line we have written the circular
Hamiltonian $L_{0}$ of the model in terms of its chiral energy-momentum tensor
$T$. A more concrete method consists in starting with a Hilbert space
representation $\ \mathcal{A}(S^{1})_{0}=\pi_{0}(\mathcal{A}(S^{1}))$ obtained
by applying the GNS representation to this vacuum state functional. This
representation has a positive energy operator given by the generator of
rotations $L_{0}$ which is quadratic in the current. It is easy to check that
the formula
\begin{align}
\left\langle W(f)\right\rangle _{\alpha}  &  :=e^{i\alpha f_{0}}\left\langle
W(f)\right\rangle _{0}\\
\pi_{\alpha}(W(f))  &  =e^{i\alpha f_{0}}\pi_{0}(W(f))\nonumber
\end{align}
defines an inequivalent state i.e. one whose GNS representation for
$\alpha\neq0$ is unitarily inequivalent to the vacuum representation with
positive energy $\left(  L_{0}\right)  _{\alpha}\equiv\pi_{\alpha}(L_{0}).$
The particular realization of the GNS representation of the $\alpha$-state in
the second line is economical because in this way the inequivalent description
becomes incorporated into the Hilbert space of the vacuum representation. For
certain generalizations in the next section it is convenient to rephrase this
result as the result of two steps, first a definition of an automorphism
$\gamma_{\alpha}$ on the $C^{\ast}$-Weyl algebra $\mathcal{A}(S^{1})$ and then
the subsequent application of the vacuum state \cite{B-M-T}
\begin{align}
&  \left\langle W(f)\right\rangle _{\alpha}=\left\langle \gamma_{\alpha
}(W(f))\right\rangle _{0},\text{ }\gamma_{\alpha}(W(f)):=e^{i\alpha f_{0}%
}W(f)\\
&  \gamma_{\alpha}(W(f))=\Gamma_{\alpha}W(f)\Gamma_{\alpha}^{\ast}%
,\,\Gamma_{\alpha}\Omega=\Omega_{\alpha},\,\left\langle .\right\rangle
_{\alpha}=\left(  \Omega_{\alpha},.\Omega_{\alpha}\right) \nonumber
\end{align}
where in the second line the automorphism is implemented by a charge-carrying
operator $\Gamma_{\alpha}$ which intertwines between the vacuum Hilbert space
$H_{0}$ and the use of the same Hilbert space for the charged representation
denoted by $H_{\alpha}$ (in order to indicate its different use)$.$ The
charge-transfer operator $\Gamma_{\alpha}$ interwines the various copies of
identical Hilbert spaces $H_{\alpha}$ and in particular relates the vacuum
state to the ground state $\Omega_{\alpha}$ of the new sector (by definition
$\Gamma_{\alpha}$ creates a rotational homogeneous charge distribution (i.e. a
distribution without radial excitations) so that the full (in this case
inseparable) Hilbert space becomes the direct (orthogonal) sum $H_{full}%
=\oplus H_{\alpha}$. Arbitrary charge distributions $\rho_{\alpha}$ of total
charge $\alpha$ i.e. $\rho_{\alpha}\left[  1\right]  \equiv\int\frac{dz}{2\pi
i}\rho_{\alpha}=\alpha$ are obtained in the form
\begin{equation}
\psi_{\rho_{\alpha}}^{\zeta}=\kappa(\rho_{\alpha})W(\hat{\rho}_{\alpha}%
^{\zeta})\Gamma_{\alpha} \label{form}%
\end{equation}
where $\kappa(\rho_{\alpha})$ is a phase factor and the net effect of the test
function in the Weyl operator is to modify the homogeneous charge distribution
created by $\Gamma_{\alpha}$ in order to obtain $\rho_{\alpha}$ with the same
total charge \cite{B-M-T}. The necessary charge-neutral compensating test
function $\hat{\rho}_{\alpha}^{\zeta}$ is uniquely determined in terms of
$\rho_{\alpha}$ apart from a choice of one point $\zeta\in S^{1}$(the
determining equation involves the $lnz$ function which needs the specification
of a branch cut).

\begin{theorem}
The charge-carrying fields\footnote{It is costumary in the algebraic setting
to use the word ``field'' for operators which (in contrast to the neutral
observables) carry superselected charges and add the word pointlike if one is
referring to its traditional use.} for disjoint charge supports (supp$\rho
_{\alpha}\perp$ supp$\rho_{\beta}$) fulfill abelian braid group commutation
relation and additive fusion laws
\begin{align}
\psi_{\rho_{\alpha}}^{\zeta}\psi_{\rho_{\beta}}^{\zeta}  &  =e^{\pm i\pi
\alpha\beta}\psi_{\rho_{\beta}}^{\zeta}\psi_{\rho_{\alpha}}^{\zeta}%
\label{rel}\\
\psi_{\rho_{\alpha}}^{\zeta}\psi_{\rho_{\beta}}^{\zeta}  &  =e^{\pm i\frac
{\pi}{2}\alpha\beta}\psi_{\rho_{\alpha}+\rho_{\beta}}^{\zeta}\nonumber
\end{align}
where the $\pm$ signs depend on whether the path from supp$\rho_{\rho_{a}}$ to
supp$\rho_{\beta}$ taken in positive (counterclockwise) direction crosses
$\zeta$ or not. A change of the cut $\zeta$ leads to the appearance of a
charge factor
\begin{equation}
\psi_{\rho_{\alpha}}^{\zeta_{1}}\left(  \psi_{\rho_{\alpha}}^{\zeta_{2}%
}\right)  ^{\ast}=e^{\pm i\pi\alpha\beta}e^{2\pi iQ\alpha} \label{charge}%
\end{equation}
where the charge operator $Q$ is conjugate to $\Gamma$ in the sense.
$QH_{a}=\alpha H_{\alpha}$ or $Q\Gamma_{\alpha}=\alpha\Gamma_{\alpha}Q$ i.e.
is part of the zero mode structure.
\end{theorem}

The second relation expresses the abelian fusion law of the model. Up to now
the Hilbert space was the nonseparable Hilbert space of all charges and in
order to get away from this unrealistic feature of our toy model we search for
an argument which leads to charge quantization in a natural manner. It turns
out that algebraic extension of the Weyl algebra which maintain commutativity
for disjoint charge supports combined with a compatible restriction of the
inseparable Hilbert space do the job
\begin{align}
\mathcal{A}_{N}  &  =\cup_{I}\mathcal{A}_{N}(I),\mathcal{A}_{N}(I)=alg\left\{
\psi_{\rho_{\alpha}}^{\zeta}|\,supp\rho_{\alpha_{gen}}\in I,\alpha_{gen}%
=\sqrt{2N}\right\}  |_{H_{res}}\\
H_{res}  &  =\left\{  \Psi\in H\;|\,e^{2\pi iQ\alpha}\Psi=\Psi\right\}
,\,H_{res}=\oplus_{n=0}^{2N-1}H_{n}\nonumber
\end{align}
Clearly the vacuum space of the extended algebra $\mathcal{A}_{N}$ contains
all integer multiples of the old locality-preserving generating charge
$\alpha=\alpha_{gen}\mathbb{Z}$ (the charge neutral $\psi_{\rho_{\alpha}}%
\psi_{\rho_{\alpha}^{\prime}}^{\ast}$ products lead back to the original Weyl
algebra). The restricted Hilbert space $H_{res}$ is a orthogonal sum of new
charges $Q=\frac{1}{\sqrt{2N}}\mathbb{Z}\,/\alpha_{gen}\mathbb{Z\simeq Z}%
_{2N}$ $\ $i.e. consists of the dual (to the old) charge spectrum $\frac
{1}{\sqrt{2N}}\mathbb{Z}\,.$ (which has been ``neutralized''). The effect of
the mod counting is that the old charges are neutralized by enlarging the
algebra (always with its local net structure) from $\mathcal{A}$ to
$\mathcal{A}_{N},$ so that the superselection structure becomes finite (the
model becomes ``rational''). The charge-carrying fields in the new setting are
also of the above form (\ref{form}) but now the generating field carries the
charge $\int\frac{dz}{2\pi i}\rho_{gen}=Q_{gen}$ which is a $\frac{1}{2N}$
fraction of the old $\alpha_{gen}.$ Their commutation relations for disjoint
charge supports are ``braidal'' (or ``plektonic''\footnote{In the abelian case
the terminology ``anyonic'' enjoys widespread popularity, but in the present
context the ``any'' does not go well with the present emphasis on charge
quantization.} which sounds more in par with bosonic/fermionic). These objects
considered as operators localized on $S^{1}$ do depend on the cut $\zeta,$ but
using an appropriate finite covering of $S^{1}$ this dependence is removed. So
the field algebra $\mathcal{F}_{\mathbb{Z}_{2N}}$ (as opposed to the bosonic
observable algebra $\mathcal{A}_{N}$) which they generate has its unique
localization structure on a finite covering of $S^{1}.\,$An equivalent
description which gets rid of $\zeta$ consists in dealing with operator-valued
sections on $S^{1}.$

In abelian current algebras the transition from bounded Weyl-like operators to
generating operator-valued distributional fields is especially simple; one
just approaches the $\delta$-function charge distribution (``blip'') at the
origin $\alpha\delta(z-1)$ by a sequence of smooth functions $\rho_{\alpha
}(z)$ and checks the existence of the limit%
\begin{align}
&  \Phi_{\alpha}(\varphi)=lim_{\rho_{\alpha}\rightarrow\alpha\delta
(z-1)}R_{\rho_{\alpha}}Ade^{i\varphi L_{0,\alpha}}\psi_{\rho_{\alpha}}%
^{\zeta=-1}\label{blip}\\
&  \left(  \Gamma_{\alpha}\Omega,R_{\rho_{\alpha}}Ade^{i\varphi L_{0,\alpha}%
}\psi_{\rho_{\alpha}}^{\zeta=-1}\Omega\right)  =1\nonumber
\end{align}
in words: the operators $\psi_{\rho_{\alpha}}^{\zeta=-1}$ which generate
charged states with charges around the origin $\varphi=0$ are translated so
that their charge is concentrated around the angle $\varphi$ whereupon their
charge distribution is compressed to a blip at $\varphi$ in such a way that
their normalization (second line in (\ref{blip})) is maintained (which leads
to a formula for the renormalization factor $R_{\rho_{\alpha}}).$ It is
costumary to interpolate the $\delta$-function by a scaling limit
$\lambda\searrow0$ in which case the renormalization factor diverges with an
inverse power related to the scale dimension of the resulting pointlike field
\cite{B-M-T}.

The extension $\mathcal{A}\rightarrow\mathcal{A}_{N}$ which led to a
``rational'' (= finite number of sectors) charge superselection structure is a
charge-quantization extension. Most other chiral models (next section) already
come with a discrete charge spectrum. In both cases one can ask whether a
model with discrete charge superselection spectrum allows (further) local
extensions. For the abelian case at hand this would require the presence of
another generating field of the same kind as above which belongs to an integer
$N^{\prime}$ and is relatively local to the first one. This is always possible
if $N$ is divisible by a square, in fact the algebra $\mathcal{A}_{N}$ is
maximal precisely if $N$ is of the form $N=p_{1}...p_{k}$ where $p_{i}$ are
prime numbers. Whereas in abelian current models such question can be answered
in terms of pedestrian computations, the generic case is conceptually much
more challenging. In the next section we will return to these problems in a
more general setting.

Before we pass to the issue of conformal invariance and the problems of
general chiral models, we cannot resist to mention a simple yet somewhat
surprising relation between the Schwinger model (QED$_{2}$ with massless
Fermions), whose charges are screened, and the Jordan model, which has
(liberated) charge sectors. Since the Lagrangian formulation of the Schwinger
model is a gauge theory, the analog of the 4-dim. asymptotic freedom wisdom
would suggest the possibility of charge liberation in the short distance limit
of this model. This seems to contradict the statement that the intrinsic
content of the Schwinger model after removing a classical degree of
freedom\footnote{In its original gauge theoretical form the Schwinger model
has an infinite vacuum degeneracy. The removal of this degeneracy (restoration
of the cluster property) with the help of the ``$\theta$-angle formalism''
leaves a massive free Bose field (the Schwinger-Higgs mechanism). As expected
in d=1+1 the model only possesses this one phase, a characteristic feature of
all two-dimensional non-lattice models.} is the QFT of a free massive Bose
field because such a simple free field is at first sight not expected to
contain subtle informations about asymptotic charge liberation. But the
massless limit as the potential of the free abelian current really does
contain this information i.e. the observable part of the Schwinger model
passes to the charge-liberated massless Jordan model as one can demonstrate in
detail in the short distance limit \cite{Buch}. This statement is truly
intrinsic since it refers to the screened phase, unlike the 4-dimensional
asymptotic freedom statement which is based on the perturbative phase instead
of the physical quark confinement phase (the asymptotic freedom statement is
the result of a consistency check falling short of a mathematical theorem).
There are other properly renormalizable (i.e. not superrenormalizable as the
Schwinger model) two-dimensional models in which one can prove the validity of
asymptotic freedom in the physical phase, but the poor state of
nonperturbative knowledge in d=1+3 is hampering an understanding of this issue
in realistic cases beyond the level of a plausibility statement. These
considerations show in addition that there is nothing intrinsic about a gauge
theoretical formulation; in fact the gauge idea in the setting of quantum
field idea is a computational device and not a physical principle since at
least for selfinteracting vector mesons (``nonabelian gauge theory'')
renormalizability requirement only admits one perturbative model (the
appearance of other physical (Higgs) degrees of freedom follows from
consistency of perturbation theory) and where the situation already has a
unique answer from the implementation of a quantum principle
(renormalizability) no additional principle is needed. The gauge principle
rather selects between several consistent classical field theories involving
vector fields and follows from quantum renormalizability via quasiclassical
approximations. The non-intrinsic nature and the absence of a quantum gauge
principle is also implicit in many conjectures where a gauge theoretic
formulation is expected to be dual to a non-gauge theory.

As a result of the peculiar nature of the zero mass limit of the derivative of
the massive free field, Jordan's model is also closely related to the massless
Thirring model (and the closely related Luttinger model for an interacting
one-dimensional electron gas) whose massive version is in the class of
factorizing models (see later section)\footnote{Another structural consequence
of this peculiarity leads to Coleman's theorem \cite{Co2} which connects the
Mermin-Wagner No-Go theorem for two-dimensional spontaneous continuous
symmetry breaking with these zero mass peculiarities.}. Together with the
massive version of the L-I QFT it shows two new (interconnected) properties
which are characteristic for massive d=1+1 models: the absence of an intrinsic
meaning of statistics \cite{schizon} and the emergence of a disorder variable
with a nonvanishing vacuum expectation value (disorder condensation).

The Thirring model proper is a special case in a large class of
``generalized'' (multi-coupling) multi-component Thirring models i.e.
4-Fermion interactions. Under this name they were studied in the early 70s
\cite{Frish}\cite{Weiss}\cite{1974} with the particular aim to identify
massless subtheories for which the currents have a chiral decomposition and
form current algebras.

It is interesting to look in more detail at the massive version of the
Thirring model. The counterpart of the potential of the conserved Dirac
current is the Sine-Gordon field, i.e. a composite field which in the
attractive regime of the Thirring coupling obeys the so-called Sine-Gordon
equation of motion. Coleman gave an argument \cite{Co1} which however does not
reveal the limitation in the size of the coupling\footnote{The current
potential of the free massive Dirac Fermion (g=0) does not obey the
Sine-Gordon equation \cite{Sc-Tr}.}. A rigorous confirmation of the existence
of a coupling range for Coleman's equivalence was recently given in the
bootstrap-formfactor setting \cite{Ba-Ka}. Two-dimensional massive models
which have a continuous or discrete internal symmetry have ``disorder'' fields
which are local fields (with respect to themselves) which implement a
``half-space'' symmetry on the charge-carrying field (acting as the identity
in the other half axis). These are pointlike bosonic fields which live in the
same Hilbert space as the charged fields, but only create the neutral part
from the vacuum. Multiplication of disorder fields with the defining field of
the model (leading short distance part of the product of the disorder operator
with the charged field) generates ``order'' fields which act cyclically on the
vacuum. The order/disorder fields have an interesting connection with phase
transitions. Whereas in the lattice version the correlation functions
\cite{Mc-Wu} of the L-I model the system undergoes a second order phase
transition as the temperature passes through the critical value, the mass
parameter represents only the slope of temperature at criticality and lost its
role of analytically connecting two phases; the only memory of the different
phases in the QFT resulting in the scaling limit consists in the presence of a
pair of order/disorder variables whose interchange in passing from one phase
to another has to be decreed as an additional rule. The resulting n-point
order/disorder correlation functions of the L-I field theory can be
represented in terms of order/disorder variables of a free Majorana field or
as the (suitable defined) square root of the exponential disorder field of a
free Dirac Fermion, both in the massive \cite{Sc-Tr} as well as in the
massless limit \cite{Re-Sc}\cite{Re1}. They are scalar Bose fields with a
$Z_{2}$ ``half-space'' commutation relation between them. Whereas the massive
scaling limit fields still have correlation functions which are order/disorder
unsymmetrical, the conformal invariant zero mass limit leads to a symmetric
situation where both variables carry superselected charges. The emergence of
new charges in connection with the appearance of critical exponents of
order/disorder fields in 2-dim. QFT is actually the content of a general
theorem \cite{Ko-Ma}.

\section{The general conformal setting and chiral theories}

Chiral theories play a special role within the setting of conformal quantum
fields. General conformal theories have observable algebras which live on
compactified Minkowski space (S$^{1}$ in the case of chiral models) and
fulfill the Huygens principle, which in an even number of spacetime dimension
means that the commutator is only nonvanishing for lightlike separation of the
fields. The fact that this rule breaks down for non-observable ``would be''
conformal fields (e.g. the massless Thirring field) was noticed at the
beginning of the 70s and considered paradoxical at that time
(``reverberation'' in the timelike (Huygens) region). Its resolution led
1974/75 to two differently but basically equivalent concepts about globally
causal objects. They are connected by the following global decomposition formula%

\begin{equation}
A(x_{cov})=\sum A_{\alpha,\beta}(x),\,\,A_{\alpha,\beta}(x)=P_{\alpha
}A(x)P_{\beta},\,\,Z=\sum e^{id_{\alpha}}P_{\alpha} \label{decomp}%
\end{equation}

On the left hand side the field lives on the universal covering of the
conformal compactified Minkowski space $\tilde{M}$. These are the
Luescher-Mack fields \cite{Lu-Ma} which ``live'' in the sense of quantum (=
modular) localization on the universal covering spacetime (or a finite
covering, depending on the model) and fulfill the global causality condition
discovered by I. Segal \cite{Se}. In the presence of interactions they are
highly reducible under the center of the covering group. The objects on the
right hand side are the component fields which were introduced in \cite{Sc-Sw}
with the aim to have objects which live on the projection $x(x_{cov})$ i.e. on
the (Dirac-Weyl compactified) Minkowski spacetime $\bar{M}$ of the laboratory
instead of the ``hells and heavens'' of the covering; The connection is given
by a decomposition formula into irreducible conformal blocks with respect to
the center $\mathbf{Z}$ of the covering group $\widetilde{SO(2,n)}$ where
$\alpha,\beta$ are labels for the eigenspaces of the generating unitary $Z$ of
the abelian center $\mathbf{Z.}$ Under central transformations these component
fields transform with a numerical phase which is proportional to the
difference of anomalous dimensions $d_{\alpha}-d_{\beta\text{ }}$\cite{Sc-Sw}
whereas the globally causal Luescher-Mack fields pick up an operator-valued
phase. The composition formula is minimal in the sense that in general there
will be a refinement due to the presence of additional charge superselection
rules (and internal group symmetries) which have no bearing on the covering
aspect of the Luescher-Mack fields and their algebraic commutation relations.
Technically speaking the $A_{\alpha,\beta}(x)$ are operator-valued
distributional sections (``conformal blocks'') in the compactification of
ordinary Minkowski spacetime. They are not Wightman fields since they
annihilate the vacuum if the right hand projector $P_{\beta}$ differs from the
projector onto the vacuum sector.

Note that the Huygens (timelike) region in Minkowski spacetime has an ordering
structure $x\prec y$ or $x\succ y$ (earlier, later). In d=1+1 the topology
allows in addition a spacelike left-right ordering $x\lessgtr y.$ This
together with the factorization of the group $\widetilde{SO(2,2)}%
\simeq\widetilde{PSL(2R)_{l}}\otimes\widetilde{PSL(2,R)_{r}}$ in particular
$\mathbb{Z=Z}_{l}\otimes\mathbb{Z}_{r}$ suggests a tensor factorization into
chiral components and led to an extremely rich and successful construction
program of two-dimensional conformal QFT as a two-step process: the
classification of chiral theories on the lightray and the amalgamation of
left-right chiral theories to two-dimensional local conformal QFT. The action
of the covering chiral group on the lightray coordinates is through
fractionally acting $PSL(2,R)$ Moebius transformations
\begin{align}
x  &  \rightarrow g(x)=\frac{ax+b}{cx+d},\,g=\left(
\begin{array}
[c]{cc}%
a & b\\
c & c
\end{array}
\right)  \in SL(2,R),\,i.e.\,detg=1\\
z  &  \rightarrow g(z)=\frac{\alpha z+\beta}{\bar{\beta}z+\bar{\alpha}%
},\,g=\left(
\begin{array}
[c]{cc}%
\alpha & \beta\\
\bar{\beta} & \bar{\alpha}%
\end{array}
\right)  \in SU(1,1),\,i.e.\,\left|  \alpha\right|  ^{2}-\left|  \beta\right|
^{2}=1
\end{align}
where the linear and circular descriptions are related through the Cayley
transformation $z=\frac{1+ix}{1-ix}.$

The presence of an ordering structure permits the appearance of more general
commutation relations for the above $A_{\alpha\beta}$ component fields namely%

\begin{equation}
A_{\alpha,\beta}(x)B_{\beta,\gamma}(y)=\sum_{\beta^{\prime}}R_{\beta
,\beta^{\prime}}^{\alpha,\gamma}B_{\alpha,\beta^{\prime}}(y)A_{\beta^{\prime
},\gamma}(x),\,\,x>y
\end{equation}
with numerical $R-$coefficients which (as a result of associativity and
relative commutativity with respect to observable fields) represent the Artin
braid group. Indeed, the DHR method to interpret charged fields as
charge-superselection carriers tied by local representation theory to the
bosonic local structure of observable algebras leads precisely to such a
plektonic setting. With an appropriately formulated adjustment to observables
fulfilling the Huygens commutativity, this could also be a possibility for the
higher dimensional timelike structure. But whereas the plektonic lightlike
structure is the only spacetime commutation imposition for chiral theories, a
would-be timelike plektonic structure in higher dimensions has to coexist with
the spacelike bosonic/fermionic statistics structure which appears to lead to
a much more difficult point of departure for classifications\&constructions
than in the chiral case where a wealth of models with R-coefficients of their
charge-carrying fields have been found. In the latter case the availability of
infinite dimensional loop group and Diff(S) symmetries and the connection of
the latter to the presence of the chiral stress-energy tensor give a rich
supply of Lie-field generated observable algebras on which one has constructed
the representations of the superselected charge sectors. In higher spacetime
dimensions no such geometric infinite dimensional extensions of the finite
dimensional global conformal vacuum symmetry is known. It has been observed
that the Huygens principle in conjunctions with conformal invariance leads to
quite strong restrictions on Wightman functions \cite{Ni-To}\cite{N-R-T} which
could help in the classification program and even suggest associated new
algebraic structures. On the other hand the holographic lightfront projection
leads to transversely extended chiral models which, although not quite as
simple as chiral theories themselves, seem to be more susceptible to be
analysed in terms of algebraic commutation structures (section 8).

Whereas the conformal block picture i.e. the objects on the right hand side of
(\ref{decomp}) naturally fits into a DHR approach in which one starts with a
model of observable algebras on (compactified) Minkowski spacetime and
constructs the so-called reduced field bundle (exchange algebra field
sections), the globally causal objects in the left hand side (\ref{decomp})
which are localized (in the sense of modular theory) on the covering spacetime
suggests another approach which is more in the intrinsic spirit of group
theoretical Wigner's particle representation setting. The guiding idea would
be that the modular localization concept formulated globally on an appropriate
n-sheeted covering space $\tilde{M}^{(n)}$ (for rational theories) within a
representation theoretical setting could directly lead to global objects
without going through the DHR analysis\footnote{The DHR dichotomy between
local observables and charge-carrying fields solves an important conceptual
problem, but is not necessarily useful for model constructions; e.g.
charge-carrying free fields are much simpler mathematical objects than their
associated neutral observable algebras. All the known approximation schemes
aim at the direct construction of field correlations.}. The latter has not
been formulated on coverings and this deficiency probably leads to a somewhat
artificial (in the sense of non-intrinsic) separation into outer (spacetime)
and inner symmetries (see remarks after theorem at end of this section).
Unfortunately a direct access to globally causal Luescher-Mack fields without
going through the right hand side in (\ref{decomp}) does not (yet?) exist.

From this presentation of the development of ideas about conformal QFT and
chiral models in particular one may have obtained the impression that there is
a straight line from the decomposition theory of the early 70s to the
construction of interesting two-dimensional models, however this is not the
way history of constructions of conformal QFTs developed\footnote{The algebra
of chiral energy-stress tensors was known since the early 70s, first for the
free Dirac field and subsequently as a general structural result \cite{1974}.
A later ($\sim$1976) manuscript of Luescher and Mack contained in addition to
this result the beginnings of the c-quantization (the Ising case) but this
project remained unfortunately unfinished and unpublished, see also
\cite{F-S-T}. All these early results were superseded by the 1984 work in
\cite{F-Q-S}.}. The only examples known up to the appearance of the seminal
Belavin-Polyakov-Zamolodchikov work (BPZ) \cite{B-P-Z} were the abelian
current models of the previous subsection. The floodgates of conformal QFT
were not opened by knowing an abstract setting of conformal block
decomposition but by the BPZ discovery of ``minimal models'' and their
connections to the already existing mathematics of Witt-Virasoro and Kac-Moody
algebras. A crucial step in the understanding of the minimal models, in
particular in what sense they are ``minimal'', was contained in a prior paper
of Friedan, Qiu and Shenker \cite{F-Q-S}. That paper also showed for the first
time that the positive energy representation category of the observable
algebra generated by the energy-stress tensor cannot be encoded into a
symmetry group. The FQS proof uses the setting of Verma module
representations, but it is also possible to obtain the same conclusions from a
standard field theoretic Hilbert space setting \cite{R-S}. The combinatorial
aspects of this new structures abstracted from these model observations were
axiomatized in \cite{Mo-Se} and the origin of the Artin braid group structure
as a new manifestation of Einstein causality in chiral field theory which led
to ``exchange algebras'' was analyzed in \cite{Artin}; in fact part of the
motivation behind it was to connect the post BPZ development to what was known
about conformal theories in the 70s \cite{Sc-Sw}. This was followed by a
systematic application of the DHR concepts to this new setting in
\cite{F-R-S2}\cite{Fro-Gab}.

There was an interesting, independent and much older idea for constructing
models via representing new algebraic structures which in the course of time
merged with the chiral conformal constructions. It goes back to a 1961 paper
by O. W. Greenberg \cite{Green} who proposed to construct nontrivial examples
of Wightman field theories instead of quantizing nonlinear field equations by
studying ``Lie fields'' i.e. sets of local fields $A_{i}(x)$ fulfilling the
``Lie relation'' (for simplicity for a set of Lorentz-scalar fields)
\begin{equation}
\left[  A_{i}(x),A_{j}(y)\right]  =c-number+\int C_{ij}^{k}(x,y,z)A_{k}%
(z)d^{n}z \label{Lie}%
\end{equation}
In the days of axiomatic quantum field theory this subject led to several
papers with inconclusive results \cite{Lie}. The non-abelian chiral current
algebras at the beginning of the 70s gave some obvious illustrations of this
structure, but the more interesting case was that of the generic chiral
stress-energy tensor which was proposed in \cite{1974} as illustration of a
Lie field which closes upon itself. Of course in some general sense all chiral
fields are Lie fields since the lightray locality only permits $\delta
$-functions with a finite number of derivative multiplied with pointlike
(composite) field generators of the same operator algebra, but here this
terminology refers to the existence of a finite distinguished set of
generators which close among themselves under commutation. The net result of
this research is contained in a paper by Baumann \cite{Bau} who proved that
there are no nontrivial scalar Lie fields in higher spacetime dimensions i.e.
$C_{ij}^{k}(x,y,z)\equiv0$. Similar conclusions probably hold for
tensor/spinor fields but there seems to be no proof. Examples of conformal Lie
fields are the chiral current algebras and some of the so-called W-algebras
(generalizations of the stress-energy algebra which contain additional fields
without internal symmetry group multiplicities). Since massive Lie fields do
not lead to scatteriong (not even in d=1+1), the interest in them within
two-dimensional QFT is entirely limited to chiral models. Indeed we will see
in section 5 that solvable (factorizable) massive two-dimensional theories are
characterized by a very different algebraic structure. The Lie field structure
of chiral current algebra is generally lost by processing these current
algebras through reduced tensor products, orbifold constructions, coset
constructions and (Longo-Rehren) extensions into other models, but it seems
that all known models originate by such procedures from Lie field models.

In order to not get lost in the impressive wealth of detailed knowledge about
chiral models and the associated two-dimensional conformal QFT, but also to
avoid the opposite extreme of bothering the reader with too many conceptual
generalities, I will try to keep a middle ground by presenting some salient
points in connection with two families of models which illustrate some
important structural points in concrete and pedestrian terms.

Let us start with a family which generalizes the abelian model of the previous
section. Instead of a one-component abelian current we now take n independent
copies. The resulting multi-component Weyl algebra has the previous form
except that the current is n-component and the real function space underlying
the Weyl algebra consists of functions with values in an n-component real
vector space $f\in LV$ with the standard Euclidean inner product denoted by
$(,)$. The maximal local extension now leads to $\left(  \alpha,\beta\right)
\in2\mathbb{Z}$ i.e. an even integer lattice $\mathcal{L}$ in $V,$ whereas the
Hilbert subspace $H_{L^{\ast}}$ which ensures $\zeta$-independence is
associated with the dual lattice $L^{\ast}:$ $\left(  \lambda_{i},\alpha
_{k}\right)  =\delta_{ik}$ \cite{Sta}$.$ The resulting superselection
structure (i.e. the $Q-$spectrum) corresponds to the finite group $L^{\ast
}/L.$ It offers the possibility of selfdual lattices $L^{\ast}=L$ i.e.
two-dimensional QFT whose observables have no additional representations; a
situation which only can occur in vector spaces V whose dimension is a
multiple of 8 (the most famous case is the Leech lattice $\Lambda_{24}$ in
$dimV=24$ also called the ``mooshine'' model \cite{Lep}). The observation that
the root lattices of the Lie algebras of type $A,B$ or $E$ (example su(n)
corresponding to $A_{n-1}$) also appear, suggests that the nonabelian current
algebras associated to those Lie algebras can also be implemented. This turns
out to be indeed true as far as the level 1 representations are concerned
which brings us to the next family: the nonabelian current algebras of the
mentioned Lie algebras of level k which are characterized by the commutation
relation
\begin{equation}
\left[  J_{\alpha}(z),J_{\beta}(z^{\prime})\right]  =if_{\alpha\beta}^{\gamma
}j_{\gamma}(z)\delta(z-z^{\prime})-\frac{1}{2}kg_{\alpha\beta}\delta^{\prime
}(z-z^{\prime})
\end{equation}
where $f$ are the structure constants of the underlying Lie algebra, $g$ is
the Cartan-Killing form and $k$, the level of the algebra, must be integer in
order that the current algebra can be globalized to a loop group algebra. The
Fourier decomposition of this current algebra leads to the so called affine
Lie algebras, a special family of Kac-Moody algebras. For k=1 this algebra can
be constructed as bilinears starting from a multi-component chiral Dirac
field; in addition there exists the mentioned possibility to construct it
within the previous setting of abelian algebras by extending these algebras
with certain charge-carrying (``vertex'' algebra) operators. Level k
representations can be constructed from tensor products of k level one
currents by field theoretic reductions or directly by studying the
representation theory of infinite-dimensional Kac-Moody Lie
algebras\footnote{The global exponentiated algebras (the analogs to the Weyl
algebra) are called loop group algebras.}. Either way one finds that e.g. the
SU(2) current algebra of level k has (together with the vacuum sector) k+1
sectors (inequivalent representations). The labelling of the different sectors
is equivalent to the labelling of their ground states of the conformal
Hamiltonian $L_{0}.$ With a bigger group-theoretical effort one can construct
the representation sectors, the generating charge-carrying fields (primary
fields) including their R-matrices and the associated net of chiral operator
algebras (indexed by intervals on the circle) which in the $SU(2)_{k}$ would
be denoted by $\mathcal{A}_{SU(2)_{k},n},$ n=0,...k and in the general
semisimple case require a more complicated characterization (in terms of Weyl chambers).

Current algebras were introduced in the early 70s as a means to explore the
multi-component multi-coupling Thirring model, in particular to find critical
coupling values for which the model becomes conformally invariant
\cite{Frish}\cite{Weiss}\cite{1974} and could have interesting applications in
the field theoretic treatment of critical phenomena. The question whether at
such conformal points (the prerequisite being the vanishing of beta-functions)
one can find the nonabelian analog of the Jordan bosonization received a
positive answer when Witten \cite{Witten} proposed a bosonic Lagrangian with a
topological term (which set it apart from the standard perturbative Lagrangian
quantization setting). Its name Wess-Zumino-Witten Lagrangian resulted from an
analogy of its interaction terms in its Lagrangian group-valued field
description with a 4-dimensional phenomenological Lagrangian used by Wess and Zumino.

As an important general message coming from two-dimensional solvable QFT it is
worthwhile to note that even in those cases where the model permits baptizing
it in terms of Lagrangian quantization (thus preparing the ground for a
standard renormalized perturbation approach e.g. the massive Thirring model),
the model cannot be fully solved in the Lagrangian setting but requires the
algebraic approach. In the case of the WZW Lagrangian the correlation
functions of the group-valued bosonic field are computed by identifying this
field as a two-dimensional composite formed from combining the left/right
current algebras and using the prior current algebra representation methods
\cite{A-R}. These calculations also show that there is no intrinsic physical
meaning in topological aspects of Euclidean functional integral representations.

The construction of equivalence classes of irreducible positive energy
representations for the minimal models is more tricky than that of current
algebras. The algebraic structure of those models is given by the commutation
relation of the energy-momentum tensor
\begin{align}
&  \left[  T(z),T(z^{\prime})\right]  =i(T(z)+T(z^{\prime}))\delta^{\prime
}(z-z^{\prime})+\frac{ic}{24\pi}\delta^{\prime\prime\prime}(z-z^{\prime
})\label{T}\\
c  &  <1\curvearrowright c=c_{m}=1-\frac{6}{\left(  m+2\right)  (m+3)}%
,\,m=1,2....\nonumber
\end{align}
whose Fourier decomposition yields the Witt-Virasoro algebra i.e. a central
extension of the Lie algebra of the $Diff(S^{1})$. The first two coefficients
are determined by the physical role of $T(z)$ in connection with the
generation of the Moebius transformations and the undetermined parameter $c>0$
(the central extension parameter) is easily identified with the strength of
the T two-point function whose form is completely fixed by Moebius invariance
and dimT=2. Although the structure of the T-correlation functions resembles
that of free fields (in the sense that the theory is known once one has
specified the two-point function), the realization that if c%
$<$%
1 then it is necessarily quantized according to the second line in (\ref{T})
came as a surprise one decade after the Lie-field structure of the
energy-momentum tensor was unraveled (there is no such quantization for
c$\geq1$)\footnote{A similar quantization phenomenon was discovered by Vaughn
Jones in the mathematical theory of subfactors of Jones index
$<$%
4 but as far as I know the question about a possible direct relation remained
without answer.}. The admissable values for the existence of a Hilbert space
representation are the $c_{m}$ values in (\ref{T}) and the possible values for
the conformal energy (the non-negative operator $L_{0}$) are
\begin{equation}
h_{p,q}(c_{m})=\frac{\left[  \left(  m+1\right)  p-mq\right]  ^{2}-1}%
{4m(m+1)},\,\left\{
\begin{array}
[c]{c}%
1\leq p\leq m-1\\
1\leq q\leq m
\end{array}
\right.
\end{equation}
That there are really algebras representations $\mathcal{A}_{m,p,q}$ which
fill these slots can be seen by constructing such models via a SU(2)$_{k}$
current coset construction which reduces the existence problems of these
models to that of the simpler current algebras (which can be obtained by
performing reductions on tensor product of free massless SU(2) Dirac fields).
Constructing chiral models does generally not mean the explicit determination
of the Wightman functions of their generating fields but a proof of their
existence by demonstrating that these models are obtained from free fields by
a series of controllable but often involved constructive steps as reduction of
tensor products formation of orbifolds under group actions, coset
constructions controllable extensions etc. The generating fields of the models
are not obeying free field equations (are not ``on-shell''). The cases where
one can write down n-point functions of generating fields are very rare; in
the case of the minimal family this is only possible for the Ising model. The
reason for this is that by ``doubling'' the Ising model one connects to the
exponential field of the Jordan model from where one can return to the already
fairly complex chiral Ising n-point functions of order/disorder variables by
drawing a square root in a suitable way \cite{Re1}. This simplification
through doubling works also for the massive Ising field theory \cite{Sc-Tr}.
It is even possible to construct a represenration for the Ising correlations
on a 2-dimensional Euclidean lattice \cite{Mc-Wu}.

To show the power of inclusion theory for the determination of the charge
content of theory let us look at a simple illustration in the context of the
above multi-component abelian current algebra. The vacuum representation of
the corresponding Weyl algebra is generated from smooth $V$-valued real
functions on the circle modulo constant functions (i.e. with vanishing total
integral) $f\in LV_{0}$. These functions equipped with the aforementioned
complex structure generate a Hilbert space $H_{1}=\overline{LV_{0}}.$ The
$I$-localized subalgebra is generated by the subspace of $I$-supported
functions (class functions whose representing functions are constant in the
complement $I^{\prime}$)
\begin{equation}
\mathcal{A}(I):=alg\left\{  W(f)|\;f\in LV_{0},\,f=const\,\,in\,\,I^{\prime
}\right\}
\end{equation}
The geometric one-interval Haag duality $\mathcal{A}(I)^{\prime}%
=\mathcal{A}(I^{\prime})$ (the commutant algebra equals the algebra localized
in the complement) is simply a consequence of the fact that the symplectic
complement in terms of $Im(f,g)$ consists of real functions in that space
which are localized in the complement i.e. $K(I)^{\prime}=K(I^{\prime})$ in a
self-explanatory notation. The answer to the same question for a double
interval $I=I_{1}\cup I_{2}$ of non-intersecting is more tricky but can be
worked out in the same setting by a pedestrian calculation
\begin{align}
&  K(\left(  I_{1}\cup I_{2}\right)  ^{\prime})\subset K(I_{1}\cup
I_{2})^{\prime}\label{double}\\
&  \curvearrowright K(I_{1}\cup I_{2})\subset K(\left(  I_{1}\cup
I_{2}\right)  ^{\prime})^{\prime}\nonumber
\end{align}
The meaning of the left hand side is clear, these are functions which are
constant in $I_{1}\cup I_{2}$ with the same constant in the two intervals. A
bit of thinking reveals that the symplectic complement on the right hand side
consists of functions which are also constant there but now different
constants are permitted. This statement translates via the functorial relation
into a conversion of the Haag duality to an inclusion $\mathcal{A}(I_{1}\cup
I_{2})\subset\mathcal{A}(\left(  I_{1}\cup I_{2}\right)  ^{\prime})^{\prime}.$
Physically the enlargement results from the fact that within the charge
neutral vacuum algebra a charge split with one charge in $I_{1}$ and the
compensating charge in $I_{2}$ for all values of the (unquantized) charge
occurs. A more realistic picture is obtained if one allows a charge split to
begin with, but one which is controlled by a lattice $f(I_{2})-f(I_{4})\in2\pi
L$ (where $f(I)$ denotes the constant value $f$ takes in that interval)$.$
Although imposing such a lattice structure destroys the linearity of the
symplectic space underlying the Weyl algebra and hence the functorial relation
between one-particle spaces and Weyl algebras, one can nevertheless define
generalized Weyl generators which generate an operator algebra $\mathcal{A}%
_{L}(I_{1}\cup I_{2}).$ It is easy to check that $\mathcal{A}_{L^{\ast}}%
(I_{1}\cup I_{2})\supset\mathcal{A}_{L}(I_{1}\cup I_{2})$ with $L^{\ast}$
being the dual lattice (which contains the original lattice) commutes with
$\mathcal{A}_{L}(\left(  I_{1}\cup I_{2}\right)  ^{\prime}),$ but in order to
show $\mathcal{A}_{L}(\left(  I_{1}\cup I_{2}\right)  ^{\prime})^{\prime
}=\mathcal{A}_{L^{\ast}}(I_{1}\cup I_{2})$ one has to work a little bit harder
since one cannot refer the algebraic to a one-particle spatial relation when
lattices are involved. When we chose $L$ to be even integer we are back at the
previous extension situation for which the charge structure is described by
the finite group $G=L^{\ast}/L.$ In fact using the conceptual framework of
Vaughan Jones one can show that the two-interval inclusion is a Jones
inclusion which is independent of the position of the disjoint intervals
characterized by the group $G$; in particular the Jones index (a measure of
the size of the bigger in terms of the smaller algebra is the Jones index
\begin{align}
&  ind\left\{  \mathcal{A}_{L}(I_{1}\cup I_{2})\subset\mathcal{A}_{L}(\left(
I_{1}\cup I_{2}\right)  ^{\prime})^{\prime}\right\}  =\left|  G\right| \\
&  \mathcal{A}_{L}(I_{1}\cup I_{2})=inv_{G}\mathcal{A}_{L^{\ast}}(I_{1}\cup
I_{2})\nonumber
\end{align}
There exists another form of this inclusion which is more suitable for
generalizations. One starts from the quantized charge extended local algebra
$\mathcal{A}_{L}^{ext}\supset\mathcal{A}$ described before in terms of an
integer even lattice $L$ (which lives in the separable Hilbert space
$H_{L^{\ast}})$ as our observable algebra. Again the Haag duality is violated
and converted into an inclusion $\mathcal{A}_{L}^{ext}(I_{1}\cup I_{2}%
)\subset\mathcal{A}_{L}^{ext}(\left(  I_{1}\cup I_{2}\right)  ^{\prime
})^{\prime}$ which reveals the same $L^{\ast}/L$ charge structure (it is in
fact isomorphic to the previous inclusion). In the general setting (current
algebras, minimal model algebras,...) this double interval inclusion is
particularly interesting if the associated Jones index is finite. One finds \cite{K-L-M}

\begin{theorem}
A chiral theory with finite Jones index $\mu$
\begin{equation}
\mu=ind\left\{  \mathcal{A}_{ext}(\left(  I_{1}\cup I_{2}\right)
):\mathcal{A}(I_{1}\cup I_{2})\right\}
\end{equation}
for the double interval inclusion which is strongly additive and split is a
rational (finite number of superselection sectors) theory and the statistical
dimensions $d_{\rho}$ of its charge sectors are related to this Jones index
through the formula
\begin{equation}
\mu=\sum_{\rho}d_{\rho}^{2}%
\end{equation}
\end{theorem}

Instead of going further into the zoology of models it may be more revealing
to mention some of the algebraic methods by which they are constructed and
explored. The already mentioned DHR theory provides the conceptual basis for
converting the notion of positive energy representation sectors (equivalence
classes of unitary representations) of the chiral model observable algebra
$\mathcal{A}$ into endomorphisms $\rho$ of this algebra. This is an important
step because contrary to group representations which have a natural (tensor
product) composition structure, representations of operator algebras (beyond
loop groups) do not come with a natural composition. The DHR theory of
localized endomorphisms of $\mathcal{A}$ leads to fusion laws and an intrinsic
notion of generalized statistics (for chiral theories: plektonic in addition
to bosonic/fermionic). The chiral statistics parameter are complex numbers
whose phase is related to a generalized concept of spin via a spin statistics
theorem and whose absolute value (the inverse of the statistics dimension)
generalizes the notion of multiplicities of fields known from the description
of inner symmetries in higher dimensional standard QFTs. The different sectors
may be united into one bigger algebra called the exchange algebra in the
chiral context (the ``reduced field bundle'' of DHR) in which every sector
occurs with multiplicity one and the statistics data are encoded into exchange
(commutation) relations of charge-carrying operators (``exchange fields'')
\cite{Artin}\cite{F-R-S2}. Even though all properties concerning fusion and
statistics are nicely encoded into this algebra, it lacks some cherished
properties of standard field theory: there is no unique state--field relation
i.e. no Reeh-Schlieder property; if a field whose source projection does not
coalesce with the projection onto the vacuum sector hits the vacuum, it
annihilates the latter. In operator algebraic terms, the local algebras are
not factors. This poses the question of how to construct from the set of all
sectors natural extensions (not necessarily local) with these desired
properties. Despite numerous attempts using different concepts, no natural
solution to this internal symmetry problem and an associated field algebra in
analogy to the DR group symmetry \cite{D-R} was found. Even the approach based
on the concept of Quasi-Hopf quantum symmetries \cite{Ma-Scho}, which at least
seems to be wide enough to cover all rational models, lacks intrinsicness and
naturality\footnote{All attempts are post factum i.e. none of them has been
used in the construction of models. In fact they seem to be less useful than
the reduced field bundle (the exchange algebra) which at least follows simple
rules and does not create the mentioned problem.} as a result of a not very
attractive mixing of global with local aspects which causes non-localities in
the relation of the gauge invariant observables to the charge-carrying
quasi-Hopf objects. On the other hand it was found \cite{Lo-Re} that natural
extensions can be characterized in operator algebraic terms by the existence
of so called DHR triples ($\Theta,w,v$)\thinspace where the so-called dual
canonical endomorphism $\Theta$ is an endomorphism of $\mathcal{A}$ which
decomposes into the sector-associated irreducible DHR endomorphisms and $w,v$
two intertwining operators in $\mathcal{A}$ which fulfill specific interwining
relations which assure the existence of a natural extension $\mathcal{A}%
\subset\mathcal{B}_{(\Theta,w,v)}.$ But in general there is no extension which
like the DHR field algebra combines all existing sectors into one object. In
case of rational theories the number of such extensions is finite and in the
aforementioned ``classical'' current algebra- and minimal- models they all
have been constructed by this method \cite{Ka-Lo1}\cite{K-L-R}, thus
confirming and completing the previous incomplete less systematic
constructions. The same method adapted to the chiral tensor product structure
of d=1+1 conformal observables classifies and constructs all two-dimensional
local (bosonic/fermionic) conformal QFT $\mathcal{B}_{2}$ which can be
associated with the observable chiral input. It turns out that this approach
leads to another of those pivotal numerical matrices which encode structural
properties of QFT: the coupling matrix $Z$%
\begin{align}
\mathcal{A\otimes A}  &  \subset\mathcal{B}_{2}\\
\sum_{\rho\sigma}Z_{\rho,\sigma}\rho(\mathcal{A})\otimes\sigma(\mathcal{A})
&  \subset\mathcal{A\otimes A}\nonumber
\end{align}
where the second line is an inclusion solely expressed in terms of observable
algebras from which the desired (isomorphic) inclusion in the first line
follows by a canonical construction, the so-called Jones basic construction.
The numerical matrix $Z$ is closely related to the so-called statistics
character matrix and it has also a deep relation to the matrix S appearing in
the SL(2,Z) modular character transformation (see also subsection 5). However
unlike the construction of the DR field algebra, these extension methods
generally do not lead to objects which incorporate all superselection sectors.
Unlike the Luescher-Mack idea of aiming directly at globally causal fields on
the covering spaces, the Longo-Rehren extension method is purely algebraic
i.e. does not incorporate the global covering aspects in its present form.

The chiral extension problem is also closely related to the problem of
amalgamating left and right chiral representations in order to arrive at local
two-dimensional conformal algebras. Hence it is not surprising that also the
construction of all two-dimensional models associated to c%
$<$%
0 chiral models has been successfully completed by these extension ideas
\cite{Ka-Lo2}.

Whether all the different construction ideas (coset and ``orbifold''
constructions starting from known models, extensions) are sufficient for a
complete classification of chiral models is an open problem.

\section{QFT in terms of modular positioning of ``monade algebras''}

QFT has been enriched by a the powerful new concept of modular localization
which promises to revolutionize the task of (nonperturbative) classification
and construction of models. It also provides an additional strong link between
two-dimensional and higher dimensional QFT and admits a rich illustration for
chiral theories. For a description of its history and aims, the reader is
referred to \cite{Bo}\cite{Yng}\cite{Sc3}

It had been known for some time that under very general conditions (for
wedge-localized algebras and interval localized algebras of chiral QFT no
additional conditions need to be imposed) the localized operator algebras
$\mathcal{A}(\mathcal{O})$ of AQFT are isomorphic to an algebra which belongs
to a class which already appeared in the famous classification of factor
algebras by Murray and von Neumann and whose special role was highlighted
later in mathematical work by Connes and Haagerup. It is somewhat surprising
that the full richness of QFT can be encoded into the relative position of a
finite number of copies of this ``monade''\footnote{We borrow this terminology
from the mathematician (co-inventer of calculus) and philosopher Gottfried
Wilhelm Leibnitz; in addition to its intended philosophical content it has the
advantage of being much shorter than the full mathematical terminology
``hyperfinite type III$_{1}$ Murray-von Neumann factor''. Instead of ``a
finite number of copies of the (abstract) monade'', we will simply say ``a
finite number of monades''} within a common Hilbert space \cite{Ka-Wi}. Chiral
conformal field theory offers the simplest theoretical laboratory in which the
emergence of the spacetime symmetry of the vacuum (the Moebius group) and the
spacetime indexed (intervals on the compactified lightray) operator algebras
can be analyzed by starting from 3 monades in a certain relative position of
modular inclusion. A modular inclusion of two monades $\left(
\mathcal{A\subset B},\Omega\right)  $ in a joint standard situation (common
standard vector $\Omega$) has two modular groups. If the $\sigma
_{t}^{\mathcal{B}}$ acts on the smaller algebra for $t<0$ as a one-sided
compression $\sigma_{t}^{\mathcal{B}}(\mathcal{A})\subset\mathcal{A,}$ the two
unitaries $\Delta_{\mathcal{A,B}}^{it}$ modular groups generate a unitary
representation of a positive energy spacetime translation-dilation group with
the (Anosov) commutation relation
\begin{equation}
Dil(\lambda)U(a)Dil^{\ast}(\lambda)=U(\lambda a),\,Dil(e^{-2\pi t}%
)=\Delta_{\mathcal{B}}^{it}%
\end{equation}
The geometrical picture which goes with this abstract modular inclusion is
$\mathcal{B}=\mathcal{A}(I)\supset\mathcal{A}(I^{\prime})=\mathcal{A}$ with
the two intervals $I^{\prime}\subset I$ having one endpoint in common so that
the modular group of the bigger one ($\simeq$ $Dil_{I}=$ Moebius
transformation leaving $\partial I$ fixed) leaves this endpoint invariant and
compresses $I^{\prime}$ into itself by transforming the other endpoint of
$\partial I^{\prime}$ into $I^{\prime}.$ One can show that this half-sided
modular inclusion ($\pm hsm,$ $t\lessgtr0$) actually forces the von Neumann
algebras to be copies of the monade.

The simplest way to obtain the full Moebius group as a symmetry group of a
vacuum representation is to require that the modular inclusion itself is
standard\footnote{There are other equivalent algebraic assumptions about
monades (hsm factorization, modular intersection) which are more convenient
for higher dimensional generalizations of the monade generation of QFT.} which
means that in addition $\Omega$\ is also standard with respect to the relative
commutant $\mathcal{A}^{\prime}\cap\mathcal{B.}$

\begin{theorem}
\bigskip The observable algebras of chiral QFT are classified by standard hsm
of two monades.
\end{theorem}

The net of interval-indexed local observable algebras is obtained by applying
the Moebius group to the original monade $\mathcal{A}\,$or $\mathcal{B.}$

The reader may have wondered why we did not follow the classical analysis of
conformal symmetry (based on transformations which leave the Minkowski metric
invariant up to a spacetime dependent factor) which in d=1+1 leads to the
infinite diffeomorphism group. Certainly all of the afore-mentioned models
have energy-momentum tensors whose Fourier decomposition leads to the unitary
implementation of $Diff(S^{1}).$ But there are also Moebius-invariant chiral
models which do not originate from chiral decomposition of two-dimensional
conformal theories but rather from holographic projections of higher
dimensional QFT.

In passing it may be helpful to point out that most of the literature on
chiral QFT is conceptually flawed on the meaning of two-dimensional conformal
invariance and in particular about conformal invariance of chiral components.
The textbook folklore claims that the spatial symmetry is described by a
hypothetical group of ``all analytic transformations $z\rightarrow f(z)".$
This is incorrect since there is no such group (i.e. the functions which are
analytic in a region of the complex plane do not form a Lie group or Lie
algebra). The only group which describes the symmetry of the (compactified)
plane is the Moebius group. The correct group is $Diff(S^{1})=\cup
_{n=1}^{\infty}Diff_{n}(S^{1})$ (of which the Moebius group $PSU(1,1)\equiv
Diff_{1}(S^{1})$ is the only vacuum-preserving subgroup). This is the group of
\textit{quasisymmetric} transformation, a subgroup of \textit{quasiconformal}
transformations which have a finite (increasing with n) distance from the
Moebius group, where the distance is defined in a topology in terms of the
Schwartz derivative \cite{Lehto}. Related to this conceptual flaw is the very
unfortunate terminology of calling local chiral fields ``holomorphic''.The
object behind the holomorphic properties of certain chiral correlation
functions is the vacuum state; it is not a property of operators or algebras
since it is immediatly lost in other states i.e. it would have been more
sensible to consider ``holomorphic'' to be an attribute of the chiral vacuum.
In particular there are no normalizable eigenstates of $Diff_{n}(S^{1})$ for n%
$>$%
1; however, and this will be the main point of this section, there are
\textit{partially invariant modular states} which for fixed n have the same
action as $Diff_{n}(S^{1})$ but only on suitably chosen $Diff_{n}(S^{!}%
)$-invariant subalgebras.

The only known counterexamples of models which are Moebius invariant but lack
the full $Diff(S^{1})$ covariance can be excluded on the basis of two well
motivated quantum physical properties: strong additivity and the split
property \cite{Ca-We}. So the question whether with these requirements the
extension from vacuum preserving Moebius invariance to Diff(S) covariance is
guarantied is a natural one. The fact that chiral structures do not only come
from 2-dim. conformal QFT but also from holographic projections in higher
dimensional massive QFT lends importance to this question. It is easy to see
that if one assumes Diff(S) covariance then transformations $z\rightarrow
z^{\kappa},0<\kappa<1$ (an angular down-scaling) have implementing
isomorphisms between restricted algebras on whose localization region the
transformation defines an invertible diffeomorphism between intervals (e.g. on
open interval contained in S) since any such partial diffeomorphism may be
completed to a global Diff(S) whose restricted automorphic action which leads
to the isomorphism does not depend on how it was extended. Modular theory
applied to the two algebras leads to a standard unitary implementation
$U_{I}(\kappa)$ which transforms the modular invariant vacuum state $\Omega$
of the first algebra into a one-parameter family of standard vectors
$\Omega_{\kappa}$ with respect to the image algebra $\mathcal{A}(I_{k}%
)=U_{I}(\kappa)A(I)U_{I}^{\ast}(\kappa).$ The last step consists in realizing
that the modular group of ($\mathcal{A}(I_{k}),\Omega_{k}$) is geometric and
equal to the $\kappa$-transformed dilation group of the interval $I.$ Hence
the presence of an automorphic action of Diff(S) on the observable algebra
results in a host of ``partial geometric modular situations'' which in
contrast to the Moebius group only act geometrically if restricted to the
relevant subalgebras. A particular physically attractive situation is obtained
if the algebra has the split property\footnote{A sufficient condition for the
validity of the split property in chiral models is the finiteness of the
particion function $tre^{-\beta L_{0}}<\infty.$}. In that case one can find a
standard vector $\Phi$ on which the two-fold localized algebra $\mathcal{A}%
(I)\vee\mathcal{A}(J),$ with $I=(0,\frac{\pi}{2}),$ $J=(\pi,\frac{3\pi}{2})$
being the two opposite quarter circles of the first and third quadrant, yields
a partially geometric modular group which acts as Dil$_{2}(e^{-2\pi t})$ with%

\begin{align}
z  &  \rightarrow g_{2}(z)=\left(  \frac{\alpha z^{2}+\beta}{\gamma
z^{2}+\delta}\right)  ^{\frac{1}{2}}\label{Dil(2)}\\
\left(
\begin{array}
[c]{cc}%
\alpha & \beta\\
\bar{\beta} & \bar{\alpha}%
\end{array}
\right)   &  =\left(
\begin{array}
[c]{cc}%
\cosh2\pi t & -\sinh2\pi t\\
-\sinh2\pi t & \cosh2\pi t
\end{array}
\right) \nonumber
\end{align}
Clearly this $Dil_{2}$ transformation has 4 fixed points and leaves the doubly
$I\cup J$ localized algebra invariant. It consists of $z\rightarrow z^{2}$
being followed by the Moebius dilation and the inverse of the first
transformation formally written as $z\rightarrow\sqrt{z}$. The split state on
$\mathcal{A}(I)\vee\mathcal{A}(J)\simeq\mathcal{A}(I)\otimes\mathcal{A}(J)$
(this is the split isomorphism) is
\begin{equation}
\omega_{\Phi}(AB)=(\Phi,A\Phi)(\Phi,B\Phi),\text{ }A\in\mathcal{A}%
(I),B\in\mathcal{A}(J)
\end{equation}
According to the previous remarks the partial diffeomorphism $z\rightarrow
z^{2}$ permits to re-write this state as $\omega(\hat{A})\omega(\hat{B})$ i.e.
the product of vacuum expectation values of the images $\hat{A},\hat{B}%
\in\mathcal{A}(0,\pi)$ which is left invariant under the $Dil(e^{-2\pi t})$
action and returns to the original form upon the inverse partial
diffeomorphism. These remarks amount to the statement that the validity of the
KMS condition (the criterion for a group to be the modular group of a state)
with $Dil_{2}$ in the state in $\omega_{\Phi}$ is equivalent to the KMS
condition with $Dil$ in $\omega\otimes\omega,$ the last being true by the
modular property of the Moebius dilation \cite{Sc-Wi}\cite{Fa-Sc}\cite{Lo-Xu}.
Again modular theory leads to a distinguished realization of the state
$\omega_{\Phi}$ on $\mathcal{A}(I)\vee\mathcal{A}(J)$ by a vector in the
vacuum Hilbert space. The split formalism introduces an unsymmetry between the
representation of this algebra and that of its commutant which has the
consequence that the modular action restricted to the commutant is not
geometric (in order to obtain a geometric action one has to start with the
commutant and go through the same steps). Note that the doubly localized
algebras through their violation of Haag duality (which became replaced by an
inclusion) were precisely those situations which revealed the charge content
through their charge--anti-charge splitting in the previous subsection.
Whereas the modular group of such situations can be made partially geometric
by the choice of a suitable state, the modular conjugation cannot be geometric
since it must carry the informations about the charge splitting.

The strong additivity property applied to the four quarter algebras which are
fixed by the $Dil_{2}$ group is the equality $A=\vee_{i}A_{i}$ which permits
to glue together the partial automorphisms to a $Dil_{2}$ automorphism of the
global algebra. Automorphisms of global algebras are unitarily implementable
since global algebras turn out to be type I operator algebras. Clearly by
using Moebius subgroups in (\ref{Dil(2)}) with two fixed points in different
positions one generates the diffeomorphism covariance $Diff_{2}(S)$\ which is
associated to the generators $L_{\pm2},L_{0}.$ By generalizing the above
construction to higher powers in $z$ and the corresponding inverse mappings
one obtains partial modular vectors and partial isomorphisms which lead to
partial geometric automorphisms (in the previous sense) associated with
$Diff_{n}(S)$; in this way partial geometric modular theory generates
$Diff(S^{1}).$ This gives for every local algebra $A(I)$ besides the vacuum
another infinite set of partially geometric modular vectors which, different
from the $Moeb$-invariant vacuum vector, change together with the change of
the localization region. The adjective ''partially geometric'' refers to the
fact that the modular group $\sigma_{t}^{\mathcal{A}(I),\omega_{\Phi}}$
restricted to $\mathcal{A}(I)$ acts like a diffeomorphism, but unlike the
vacuum the modular state does not generate a globally geometric action. By the
split property one can extend the construction to the more natural setting of
n-fold localized states in $\mathcal{A}.$

The interesting question of whether assumptions about the existence of
partially geometric modular states and groups can be rephrased in terms of a
natural positioning of a finite number of monades in suitable joint modular
states remains open.

This kind of problem has gained importance as a result of a recent discovery
of Brunetti, Fredenhagen and Verch \cite{B-F-V} (with important prior
observations by Hollands and Wald) which permits to formulate Einstein's local
covariance which underlies classical general relativity (physical equivalence
of isometrically diffeomorphic manifolds) in the setting of curved spacetime
QFT\footnote{It assures the ``background independence'' of the algebraic
substrate and although this property by the very quantum nature of states does
not permit to maintain it for individual states it does become transferred to
the folium of a state \cite{B-F-V}.}. For the simplest case of free fields
(Weyl algebras) BFV establish the validity of this new quantum local
covariance requirement i.e. that the model which was local in the standard
Minkowski sense also fulfills local covariance in the new sense. The quantum
version of this new principle (which as mentioned contains the locality
principle underlying standard Minkowski space QFT as a special case) adapted
to the chiral setting amounts to the question whether Moebius covariant
theories under reasonable local quantum physical assumptions are
$Diff(S)$-covariant. As already mentioned this is of course the case in all
models which possess a energy-stress tensors. Since the Moebius symmetry (in
higher dimensions also the Poincar\'{e}- or conformal- symmetry) and the
construction of Moebius invariant nets of local algebras can be fully encoded
into the relative position of a finite number of monades, it would be very
satisfying indeed to extend this algebraic setting to diffeomorphisms
$Diff(S)$.

\textbf{It is hard to imagine how one can ever combine quantum theory and
gravity without problematizing and understanding these still mysterious links
between spacetime geometry, thermal properties and relative positioning of
monades in a joint Hilbert space}. Chiral theory and d=1+1 conformal QFT offer
certainly the simplest testing ground for these new ideas.

\section{Euclidean rotational chiral theory and temperature duality}

Euclidean theory associated with certain real time QFTs is a subject whose
subtle and restrictive nature has been lost in many contemporary publications
as a result of the ``banalization'' of the Wick rotation (for some pertinent
critical remarks referred to in \cite{Re2}). The mere presence of analyticity
linking real with imaginary (Euclidean) time without establishing the subtle
reflection positivity (which is necessary\ to derive the real time spacelike
commutativity as well as the Hilbert space structure) is not of much physical
use; one needs an operator algebraic understanding of the so-called Wick rotation.

The issue of understanding Euclideanization in chiral theories became
particularly pressing after it was realized that Verlinde's observation on a
deep connection between fusion rules and modular transformation properties of
characters of irreducible representations of chiral observable algebras is
best understood by making it part of a wider investigation involving angular
parametrized thermal n-point correlation functions in the superselection
sector $\rho_{\alpha}$
\begin{align}
\left\langle \Phi(\varphi_{1},..\varphi_{n})\right\rangle _{\rho_{\alpha}%
,2\pi\beta_{t}} &  :=tr_{H_{\rho_{\alpha}}}e^{-2\pi\beta_{t}\left(
L_{0}^{\rho_{\alpha}}-\frac{c}{24}\right)  }\pi_{\rho_{\alpha}}(\Phi
(\varphi_{1},..\varphi_{n}))\,\\
\Phi(\varphi_{1},..\varphi_{n}) &  =\prod_{i=1}^{n}\Phi_{i}(\varphi
_{i})\nonumber\\
\left\langle \Phi(\varphi_{1},..\varphi_{n})\right\rangle _{\rho_{\alpha}%
,2\pi\beta_{t}} &  =\left\langle \Phi(\varphi_{n}+2\pi i\beta_{t},\varphi
_{1},..\varphi_{n-1})\right\rangle _{\rho_{\alpha},2\pi\beta_{t}}\nonumber
\end{align}
i.e. the Gibbs trace at inverse temperature $\beta=2\pi\beta_{t}$ on
observable fields in the representation $\pi_{\rho_{\alpha}}.$ Gibbs states
are special KMS states (states which fulfill the analytic property in the
third line) whose zero point function is the partition functions. Such thermal
states are (in contrast to the previously used ground states)
\textit{independent on the particular localization of charges} $loc\rho
_{\alpha},$ they only depend on the equivalence class i.e. on the sector
$\left[  \rho_{\alpha}\right]  \equiv\alpha.$ These correlation
functions\footnote{The conformal invariance actually allows a generalization
to complex Gibbs parameters $\tau$ with $Im\tau=\beta$ which is however not
neede in the context of the present discussion.} fulfill the following thermal
duality relation
\begin{align}
\left\langle \Phi(\varphi_{1},..\varphi_{n})\right\rangle _{\alpha,2\pi
\beta_{t}} &  =\left(  \frac{i}{\beta_{t}}\right)  ^{a}\sum_{\gamma}%
S_{\alpha\gamma}\left\langle \Phi(\frac{i}{\beta_{t}}\varphi_{1},..\frac
{i}{\beta_{t}}\varphi_{n})\right\rangle _{\gamma,\frac{2\pi}{\beta_{t}}%
}\label{duality}\\
a &  =\sum_{i}dim\Phi_{i}\nonumber
\end{align}
where the right hand side formally is a sum over thermal expectation at the
inverse temperature $\frac{2\pi}{\beta_{t}}$ at the analytically continued
pure imaginary values scaled with the factor $\frac{1}{\beta_{t}}.$ The
multiplicative scaling factor in front which depends on the scaling dimensions
of the fields $\Phi_{i}$ is just the one which one would write if the
transformation $\varphi\rightarrow\frac{i}{\beta_{t}}\varphi$ were a conformal
transformation law. Before presenting a structural derivation of this relation
which is based on a new Euclideanization using modular operator theory it is
instructive to check this identity in the simple abelian current model of
section3 which permits a calculation of the thermal correlation function. By a
computation which is only slightly more involved than that in the appendix C
of \cite{B-M-T} one finds the following representation of the thermal Gibbs
state two-point function in the sector $l$ of $\mathbb{Z}_{2N}$ (in the
notation of (\ref{blip}))%
\begin{align}
&  \left\langle \Phi_{-\sqrt{2N}}(0)\Phi_{\sqrt{2N}}(\varphi)\right\rangle
_{l,\tau}=\Theta_{2l,2N}(\sqrt{2N}\varphi,\tau,0)\times\\
&  \times\frac{1}{\eta(\tau)}\left[  2isin\frac{1}{2}\varphi\prod_{\nu
=1}^{\infty}\frac{(1-2e^{i2\pi\tau\nu}cos\varphi+e^{4\pi i\tau\nu}%
)}{(1-e^{i2\pi\tau\nu})^{2}}\right]  ^{-\alpha^{2}}\nonumber\\
&  tr_{H_{l}}e^{i\tau\left(  L_{0}^{\alpha}-\frac{1}{24}\right)  }%
e^{i\sqrt{2N}\varphi Q}=\frac{1}{\eta(\tau)}\sum_{n\in\mathbb{Z}}e^{i\pi
\tau(n\sqrt{2N}+\frac{l}{\sqrt{2N}})^{2}+i\varphi\sqrt{2N}(n\sqrt{2N}+\frac
{l}{\sqrt{2N}})}=\nonumber\\
&  =\frac{1}{\eta(\tau)}\Theta_{2l,\sqrt{2N}}(\sqrt{2N}\varphi,\tau
,0)\nonumber
\end{align}
The first line contains the classical Jacobi theta-function and the $\eta
(\tau)$ the Dedekind eta-function. Instead of the inverse temperature
$2\pi\beta_{t}$ we have used the customary complex variable $\tau$ with
$Im\tau=\beta_{t}$ in terms of which the modular $SL(2,\mathbb{Z})$ group
covariance properties have the standard simple form\footnote{In addition to
higher-dimensional theories were locality and covariance permits to enlarge
the KMS analytic complex strip to a larger tubular region involving the
spatial coordinates \cite{Br-Bu}, chiral theories even allow to complexify the
value of the temperature $\beta.$}. The dependence on the $\mathbb{Z}_{2N}$
charge (the zero mode structure) is contained in the $\Theta$-function,
whereas the remainder is independent on the chosen $Z_{2N}$-model; in fact the
Q-dependent part of the thermal two-point function can be separated which
leads to the formula in the last two lines. The KMS property in terms of
pointlike fields together with the circular nature of $\varphi$ yields the
double periodicity in $\varphi.$ All the expressions converge for $Im\tau>0,$
and the transformation properties under $SL(2,\mathbb{Z})$ whose generators
are T: $\tau\rightarrow\tau+1$ and S: $\tau\rightarrow-\frac{1}{\tau}$ follow
from from those of $\Theta,\eta$ and the expression in the bracket. Under T
transformations the bracket is invariant whereas $\eta$ and $\Theta$ are
invariant up to a phase factor. Under S the $\Theta$ suffers a linear
transformation%
\begin{equation}
\sqrt{-i\tau}\Theta_{2l,2N}(\sqrt{2N}\varphi,\tau,0)=e^{2i\pi\frac{\sqrt
{2N}\varphi}{\tau}}\sum_{p=1-N}^{N}\frac{e^{\frac{ipl}{N}}}{\sqrt{2N}}%
\Theta_{2p,2N}(\frac{\sqrt{2N}\varphi}{\tau},-\frac{1}{\tau},0)
\end{equation}
from which one obtains the matrix S whereas the net effect of the
multiplication factors including those from the S transformation of the
bracket and the $\eta$ combine to the factor in (\ref{duality}). The
simplicity of the model permits the calculation of general n-point functions
with the result that only change in $\Theta$ consists in replacing the
$\sqrt{2N}\varphi$ by $\sqrt{2N}\sum\pm\varphi_{i}$ where the sign depends on
the sign of the charge and the number of + and - signs must be equal (charge
neutrality). The calculation can be extended to the multi-current lattice
models with the interesting possibility of encountering modular invariant
functions in case of selfdual even lattices. Since the Gibbs states are not
normalized, the character identities are actually the ``zero-point function''
part (i.e. $\Phi=1$ with $a=0)$ of the above relation namely
\begin{equation}
\chi_{\alpha}(\tau)=\sum_{\beta}S_{\alpha\beta}\chi_{\beta}(-\frac{1}{\tau})
\end{equation}
involving a matrix $S$ already appeared in section 4$.$

Under certain technical assumption within the setting of vertex
operators\footnote{The Vertex framework is based on pointlike covariant
objects, but unlike Wightman's formulation it is not operator-algebraic i.e.
the star operation is not inexorably linked to the topology of the algebra as
in $C^{\ast}$algebras of quantum mechanical origin. In addition it has no
extension to higher-dimensional theories.}, Huang recently presented a
structural proof \cite{Huang} and it seems that his assumptions in that
framework are equivalent to the standard rationality assumption (i.e. finite
number of sectors in the operator-algebraic approach). As in the case of the
above computational check, Huang's proof does not really reveal the deep local
quantum physical principles which are behind the thermal duality relation.

The fact that the character relation is a special case of a relation which
involves analytic continuation to imaginary rotational lightray coordinates
suggests that one should look for a formulation in which the rotational
Euclideanization has a well-defined operator-algebraic meaning. On the level
of operators a positive imaginary rotation is related to the Moebius
transformation $\tilde{\Delta}^{it}$ with the two fixed points $(-1,1)$ via
the formula%
\begin{equation}
e^{-2\pi\tau L_{0}}=\Delta^{\frac{1}{4}}\tilde{\Delta}^{i\tau}\Delta
^{-\frac{1}{4}}=\tilde{\Delta}_{c}^{i\tau}\label{cont}%
\end{equation}
where $\Delta^{it}$ and $\tilde{\Delta}^{it}$ represents the $SL(2,R)$ Moebius
subgroups with fixpoints $(0,\infty)$ resp. $(-1,1)$ and $\tilde{\Delta}%
_{c}^{i\tau}$ the $SU(1,1)$ subgroup with $z=(e^{-i\frac{\pi}{2}}%
,e^{i\frac{\pi}{2}})$ being fixed (the subscript $c$ denotes the compact
picture description). Note that $Ad\Delta^{\frac{1}{4}}$ acts the same way on
$\tilde{\Delta}^{i\tau}$ as the Cayley transformation $AdT_{c},$ where the
$T_{c}$ is the matrix which represents the fractional acting Cayley
transformation
\begin{equation}
T_{c}=\frac{1}{\sqrt{2}}\left(
\begin{array}
[c]{cc}%
i & 1\\
-i & 1
\end{array}
\right)
\end{equation}
Ignoring for the moment domain problems for $\Delta^{\frac{1}{4}}$(to which we
will return soon), the relation (\ref{cont}) gives an operator representation
for the analytically continued rotation with positive imaginary part $(t>0)$
in terms of a Moebius transformation with real rapidity parameter. If we were
to use this relation in the vacuum representation for products of pointlike
covariant fields $\Phi$ where the spectrum of $L_{0}$ is nonnegative, we would
with obtain with $\Phi(t)=e^{2\pi itL_{0}}\Phi(0)e^{-2\pi itL_{0}}$%
\begin{align}
\left\langle \Omega\left|  \Phi_{1}(it_{1})..\Phi_{n}(it_{n})\right|
\Omega\right\rangle _{{}}^{ang} &  =\left\langle \Omega\left|  \Phi_{1}%
(t_{1})_{c}..\Phi_{n}(t_{n})_{c}\right|  \Omega\right\rangle ^{rap}%
\label{vacuum}\\
&  =\omega_{2\pi}(\Phi_{1}(t_{1})_{c}..\Phi_{n}(t_{n})_{c})^{rap}\nonumber
\end{align}
The left hand side contains the analytically continued rotational Wightman
functions. As a result of positivity of $L_{0}$ in the vacuum representation
this continuation is possible as long as the imaginary parts remain ordered
i.e. $\infty>t_{1}>...>t_{n}>0.$ On the right hand side the fields are at
their physical boundary values parametrized with the rapidities of the compact
$\tilde{\Delta}_{c}^{it}$ Moebius subgroup of $SU(1,1)$. Note that this
rapidity interpretation implies a restriction since the rapidities associated
with $x=th\frac{t}{2}$ cover only the interval $(-1,1).$\ The notation in the
second line indicates that this is a KMS state at modular temperature
$\beta_{mod}=1$ ($\beta_{Hawking}=2\pi\beta_{mod}=2\pi$) in agreement with the
well-known fact that the restriction of the global vacuum state to the
interval (-1,1) becomes a state at fixed Hawking-Unruh temperature $2\pi.$
Note that only the physical right hand side is a Wightman \textit{distribution
in terms of a standard }$i\varepsilon$\textit{ boundary prescription}, whereas
the left hand side is an \textit{analytic function (i.e. without any boundary
prescription)}. This significant conceptual (but numerical harmless)
difference is responsible for the fact that in the process of angular
Euclideanization of chiral models\textbf{ the KMS condition\footnote{Contrary
to popular believes KMS is not equivalent to periodicity in time but it leads
to such a situation if the the involved operators commute inside the
correlation function (e.g. spacelike separated observables). } passes to a
periodicity property and vice versa}.

At this point it is helpful to look at an analogous situation within the
setting of the of the Osterwalder-Schrader Euclideanization. It is well-known
that this is the natural setting for the formulation of the Nelson-Symanzik
duality. Its thermal version is analogous to our problem. It says that a
two-dimensional QFT which obeys a KMS condition is Nelson-Symanzik dual
(interchange of space with imaginary time) to a ground state theory in a
periodic box. This of course appears almost a tautology if one in the setting
of the Feynman-Kac representation so that the Osterwalder-Schrader
Euclideanization leads to a bona fide classical statistical mechanics. In the
case one starts with a periodic quantization box (or rather interval) the use
of the Feynman-Kac presentation even suggests a stronger form the
Nelson-Symanzik symmetry: the thermal box (interval) at length $L$ and KMS
state (which is even Gibbs) at temperature $\beta$ is (generalized)
Nelson-Symanzik dual to a system in which $L$ and $\beta$ are interchanged.

Hence the analogy with the generalized Nelson-Symanzik situation suggests to
start from a rotational thermal representation in the chiral setting. For
simplicity let us first assume that our chiral theory is one of those special
selfdual lattice Weyl-like models in section 4 which have no other positive
energy representation except the vacuum in which case the statistics character
matrix is trivial i.e. $S=1$ in the above matrix relation relation
(\ref{duality}). Assume for the moment that the Gibbs temperature is the same
as the period namely $\beta_{mod}=1.$ According what was said about the
interchange of KMS with periodicity in the process of angular Euclideanization
we expect the selfdual relation
\begin{align}
&  \left\langle \Omega_{1}|\Phi(it_{1})...\Phi(it_{n})|\Omega_{1}\right\rangle
^{rot}=\left(  i\right)  ^{ndim\Phi}\left\langle \Omega_{1}^{E}\right|
\Phi^{E}(t_{1})...\Phi^{E}(t_{n})\left|  \Omega_{1}^{E}\right\rangle
^{rot}\label{an}\\
&  \left\langle \Omega_{1}|\Phi(t_{1})...\Phi(t_{n})|\Omega_{1}\right\rangle
^{rot}\equiv tr(\Omega_{1},\Phi(t_{1})...\Phi(t_{n})\Omega_{1}),\,\Omega
_{1}\equiv e^{-\pi L_{0}}\nonumber\\
&  \Phi^{E}(t_{1})^{\dagger}\equiv\tilde{J}\Phi^{E}(t_{1})\tilde{J}=\Phi
^{E}(-t_{1})^{\ast},\,\left[  \tilde{J},L_{0}\right]  =0\nonumber
\end{align}
where the analyticity according to a general theorem about thermal states
\cite{Kl-La} limits the $t^{\prime}s$ to the unit interval and requires the
ordering $1>t_{1}>...>t_{n}>0.\,$\ Thermal Gibbs states are conveniently
written in the Hilbert space inner product notation with the help of the
Hilbert-Schmidt operators $\Omega_{1}\equiv e^{-\pi L_{0}},$ in which case the
modular conjugation is the action of the Hermitian adjoint operators from the
right on $\Omega_{1}$ \cite{Haag}. Since the KMS and the periodicity match
crosswise, the only property to be checked is the positivity of the right hand
side i.e. that the correlations on the imaginary axis are distributions which
fulfill the Wightman positivity. Here the label $E$ on $\Phi(t_{1})$ denotes
the Euclideanization in the sense of the change of inner product and star
operation as presented at the end of section 2 (\ref{Eu}). For this we need
the star conjugation associated with $\tilde{J}$ which interchanges the right
with the left halfcircle which because of $L_{0}=H+\tilde{J}H\tilde{J}$
commutes with $L_{0}.$ In that case the modular group of $\Phi^{E}(t_{1}%
)=\Phi(it)$ is $e^{-2\pi tL_{0}}$ and the modular conjugation is the Ad action
of $\tilde{J}$ which changes the sign of $t$ as in the third line (\ref{an}).
Whereas the modular conjugation in the original theory maps a vector
$A\Omega_{1}$ into $\Omega_{1}A^{\ast}$ with the star being the Hermitean
conjugate, the Euclidean modular conjugation is $A^{E}\Omega_{1}%
^{E}\rightarrow\Omega_{1}^{E}\tilde{J}A^{E}\tilde{J}\equiv\Omega_{1}%
^{E}\left(  A^{E}\right)  ^{\dagger}.$ This property is at the root of the
curious selfconjugacy (\ref{an}). .

There are two changes to be taken into consideration if one passes to a more
general situation. The extension to the case where one starts with a $\beta$
Gibbs state which corresponds in the Hilbert-Schmidt setting to $\Omega
_{\beta}=e^{-\pi\beta L_{0}}$ needs a simple rescaling $t\rightarrow\frac
{1}{\beta}t$ on the Euclidean side in order to maintain the crosswise
correspondence between KMS and periodicity. Since the Euclidean KMS property
has to match the unit periodicity on the left hand side, the Euclidean
temperature must also be $\frac{1}{\beta}$ i.e. the more general temperature
duality reads%
\begin{equation}
\left\langle \Omega_{\beta}|\Phi(it_{1})...\Phi(it_{n})|\Omega_{\beta
}\right\rangle ^{rot}=\left(  \frac{i}{\beta}\right)  ^{ndim\Phi}\left\langle
\Omega_{\frac{1}{\beta}}^{E}\right|  \Phi^{E}(t_{1})...\Phi^{E}(t_{n})\left|
\Omega_{\frac{1}{\beta}}^{E}\right\rangle ^{rot}%
\end{equation}
The positivity argument through change of the star-operation remains
unaffected. This relation between expectation values of pointlike covariant
fields should not be interpreted as an identity between operator algebras. As
already hinted at the end of section 2 one only can expect a sharing of the
analytic core of two different algebras whose different star-operations lead
to different closure. In particular the above relation does not represent a
symmetry in the usual sense.

The second generalization consists in passing to generic chiral models with
more superselection sectors than just the vacuum sector. As usual the systems
of interests will be rational i.e. the number of sectors is assumed to be
finite. In that case the mere matching between KMS and periodicity does not
suffice because all sectors are periodic as well as KMS and one does not know
which sectors to match. A closer examination (at the operator level taking the
Connes cocycle properties versus charge transportation around the circle into
account) reveals that the statistics character matrix $S$ \cite{Re}\ enters as
in (\ref{duality}) as a consequence of the well-known connection between the
\textit{invariant} content (in agreement with the sector $\left[  \rho\right]
$ dependence of rotational Gibbs states) of the circular charge transport and
the statistics character matrix \cite{F-R-S2}. For those known rational models
for which Kac-Peterson characters have been computed, this matrix $S$ turns
out to be identical to the Verlinde matrix $S$ which diagonalizes the fusion
rules \cite{Verl} and which together with a diagonal phase matrix $T$
$\ $generates a unitary representation of the modular group
$SL(2,Z)\footnote{Whereas relativistic causality already leads to an extension
of the standard KMS $\beta$-strip analyticity domain to a $\beta$-tube domain
\cite{Br-Bu}, conformal invariance even permits a complex extension of the
temperature parameter to $\tau$ with $Im\tau>0.$ For this reason the chiral
theory in a thermal Gibbs state can be associated with a torus in the sense of
a Riemann surface, but note that in \textit{no physical sense} of localization
this theory lives \textit{on} a torus.}.$ Confronting the previous zero
temperature situation of angular Euclidean situation with the asymptotic limit
of the finite temperture identity, one obtains the Kac-Wakimoto relations as
an identity between the temperature zero limit and the double limit of
infinite temperature (the chaos state) and short distances on the Euclidean side.

This superselection aspect of angular Euclideanization together with the
problem in what sense this modular group $SL(2,Z)$ can be called a new
symmetry is closely related to a more profound algebraic understanding of the
relation between the analytic cores of the two algebras and requires a more
thorough treatment which we hope to return to in a separate publication.

Modular operator theory is also expected to play an important role in bridging
the still existing gap between the Cardy Euclidean boundary setting and those
in the recent real time operator algebra formulation by Longo and Rehren
\cite{Lo-Re}.

\section{``PFG'', factorizing models and generators of wedge algebras}

In contrast to conformal two-dimensional QFT the DHR where the DHR
superselection structure is important for the classification and construction
of models, the issue of statistics looses its physical relevance in massive
two-dimensional models\footnote{This can be traced back to the absence of a
compact (rotation-like \ transformation) i.e. the chiral rotation (as well as
the chiral tensor factorization) is lost in the presence of a massive
particle.}. The main reason for this unexpected and somewhat peculiar aspect
is the fact that statistics ceases to be an intrinsic attribute of
two-dimensional particles (they are rather statistical ``schizons''
\cite{schizon}) and the commutation relation between fields can be changed at
will (e.g. from bosonic to fermionic or even plektonic) by passing to other
fields in the same Fock space. It is easy to see that the Fock space of
two-dimensional Bosons can be described in terms of fermionic or even
``anyonic'' particle creation/annihilation operators \ \cite{schizon}. The
same result holds for pointlike localized massive fields in x space. Since the
argument for pointlike covariant x-space fields is less obvious it may be
helpful to give a brief indication. Starting from a free massive Dirac field
one constructs the (pseudo)potential (or field strength) of the conserved
current (as in section 3) $j_{\mu}(x)=\varepsilon_{\mu\nu}\partial^{\nu}\phi$
which is bilinear in the fermion creation/annihilation operators. It is
spacelike bosonic with respect to itself but has a spacelike step function
commutation relation relativ to the Dirac field; hence by mutiplying its
exponential with strength $\alpha$ (section 3) with the Dirac field and
Wick-ordering one obtains a pointlike field $\psi^{(\alpha)}(x)$ and its
conjugate which form a complex $\alpha$-anyonic field. In contradistiction to
the massless case where this construction would have led to a charge-carrying
field which transforms into different (orthogonal) charge sector, all $\alpha
$-anyonic fields (including a bosonic field for a special choice of $\alpha$)
act cyclically on the vacuum and generate the same Hilbert space. The
manifestation of this observation within the DHR setting is well-known
\cite{Mueger}: the (global) gauge-invariant subalgebra of field algebras with
an internal symmetry do not fulfill Haag duality, and those observable
algebras which do satisfy Haag duality cannot have a nontrivial charge
superselection structure. Although it is possible to define the observable
algebras in such a way that the different $\alpha$-values can be interpreted
as different superselection sectors, this viewpoint is not natural and in
particular it is not useful in the construction of massive models.

The $\psi^{(\alpha)}$-model illustrates another property which goes against
some popular folklore concerning the connection of conformal algebras with
their massive counterpart: there is no algebraic correspondence between
operators in the massive theory with those in their massless scaling limit;
rather the massive algebra contains many more operators which have no
conformal counterpart i.e. they converge to zero in the massless limit done
analogously as in section 3. To illustrate this ``thinning out'' phenomenon of
the scaling limit, one only needs to look at the trajectory of the
$\psi^{(\alpha)}$ as $\alpha$ increases. The short distance singularity
increases and the anyonic commutation behavior is periodic with period 1 (in
suitable normalization convention for $\phi$); if $\psi^{(\alpha_{red})}$
denote the anyonic fields in the interval $0\leq\alpha_{red}<1,$ the field for
a generic $\alpha$-value $\alpha=\alpha_{red}+n$ differs from $\psi
^{(\alpha_{red})}$ by a local operator (even degree in the free Fermions); in
more sophisticated terminology: the massive anyonic Borchers class is
generated by a reduced element and the even part of the free fermion Borchers
class. The process does not lead to the expected massless $\frac{\alpha^{2}%
}{2}$ trajectory rather the mass power matched to the expected short-distance
dimension leads to a a vanishing limit as a consequence of the appearance of
cumulative mass singularities (the same phenomenon which is responsible for
the restriction of the Sine-Gordon-Thirring model model equivalence outside a
certain range of coupling strength).

The non-intrinsic nature of statistics in d=1+1 massive QFT is related to the
new phenomenon of the appearance of order/disorder- and quantum soliton-
fields \cite{Freden}\cite{Mueger}. However the important structural aspect of
two-dimensional massive theories which has led to the construction of an
impressive wealth of model has been scattering theory of massive particles and
its connection with integrability (and not the structure of commutators
between spacelike separated fields).

The origin of these developments can be traced back to two ideas which
attracted a lot of attention during the 60s and 70s. On the one hand there was
the quite old idea to bypass the ``off-shell'' field theoretic approach to
particle physics (in particular strong interactions) in favor of a pure
on-shell S-matrix setting which, as the result of the elimination of short
distances via the mass-shell restriction would be free of ultraviolet
divergencies. This idea was enriched in the 60s by the crossing property which
in turn led to the bootstrap idea as a (highly nonlinear) selfconsistent
method for the determination of the S-matrix. The protagonists of this
S-matrix bootstrap program placed themselves in a totally antagonistic
position with respect to QFT so that the strong return of QFT in the form of
gauge theory undermined the credibility of the bootstrap approach. On the
other hand there were rather convincing quasiclassical calculations on certain
two-dimensional QFTs as the Sine-Gordon model suggesting that they were
quantum integrable systems and that in particular their quasiclassical mass
spectrum was exact \cite{Dashen}. These provocative observations asked for a
structural explanation beyond quasiclassical approximations and it became soon
clear that the natural setting was that of the fusion of boundstate poles of
unitary crossing symmetric purely elastic S-matrices; first in the special
context of the Sine-Gordon model \cite{S-T-W} and afterwards as a general
classification and computation program from which factorizing S-matrices can
be determined by solving well-defined equations for the elastic 2-particle
S-matrix \cite{K-T-T-W}. This line of research led finally to a general
program of a bootstrap-formfactor construction of so-called d=1+1 factorizable
models \cite{K-W}\cite{Zam}\cite{Smir}. This formfactor program uses the very
ambitious original S-matrix bootstrap idea in the limited context of a d=1+1
S-matrix Ansatz in which $S$ factorizes into 2-particle elastic components
$S^{(2)}$. A consequence of this simplification is that the classification and
calculation of factorizing S-matrices can be separated from the problem of the
construction of the associated off-shell QFT. Hence the S-matrix bootstrap
becomes the first step in a bootstrap-formfactor program, followed by a second
step which consists in calculating generalized formfactors of fields and
operators\footnote{The formfactors of an operator are defined as its matrix
elements between multi-particle bra out- and ket in- vectors.} beyond that of
the identity operator (whose formfacors correspond to the S-matrix entries).
Of course such a two-step approach is limited to factorizable models; for more
general models the construction of the S-matrix cannot be separated from the
general formfactor construction. Following an idea of Swieca, the irrelevance
of statistics is expected to manifest itself already on the level of the
S-matrix in form of a factorization into a dynamical and a
(rapidity-independent) ``statistical'' part. This would permit to reduce a
formfactor-bootstrap program with exotic commutation relation to one with
Boson/Fermion fields.

This nonperturbative bootstrap-formfactor approach for factorizing models
produced a steady stream of new models and it continues to be an important
innovative area of research. Our interest in exploring this approach lies in
the potential messages it contains with respect to a mass-shell based field
theoretic constructions without the ``classical crutches'' and ultraviolet
problems which characterize the Lagrangian quantization setting. In important
step in this direction would be an intrinsic and systematic understanding of
this subclass of models within the conceptual setting of QFT.

Limiting our interest to QFT with a mass gap we automatically secure the
validity of the powerful time-dependent (LSZ) scattering theory. Let us in
addition make the standard assumption that the Fock space of asymptotic
multi-particle states is equal to the total Hilbert space (asymptotic
completeness). To keep the notation simple, we imagine that we are dealing
with an interacting theory of just one kind of particle. Let $G$ be a
(generally unbounded) operator affiliated with the local algebra
$\mathcal{A}(\mathcal{O}).$ We call such $G$ a vacuum-\textbf{p}%
olarization-\textbf{f}ree \textbf{g}enerator (PFG) affiliated with
$\mathcal{A}(\mathcal{O})$ (denoted as $G\eta\mathcal{A}(\mathcal{O})$) if the
state vector $G\Omega$ (with $\Omega$ the vacuum) is a one-particle state
without any vacuum polarization admixture \cite{Sc3}. By definition such PFGs
are (unbounded) on-shell operators and it is well-known that the existence of
a \textit{subwedge-localized} PFG forces the theory to be interaction-free,
i.e. the local algebras possess free field generators. However, and this is
the surprising fact, this link between localized PFG and absence of
interactions breaks down if one admits wedge regions. In that case modular
theory guaranties the existence of wedge generators without vacuum
polarization; but only if these PFG are ``tempered'' (well-defined on a
translation invariant domain) \cite{B-B-S} they have useful properties in the
setting of time-dependent scattering theory. The restriction implied by this
additional requirement can be shown to only permit theories with a purely
elastic S-matrix; it has been known for a long time that this is only possible
in d=1+1 where such theories have been investigated since the late 70s within
the so-called S-matrix bootstrap program \cite{Ka-Wei}. In fact one can show
that the elastic two-dimensional S-matrices coming from local QFT are
necessarily described by two-particle S-matrices; all the higher elastic
contributions factorize into two-particle contributions and the latter are
classified by solving equations for functions in the rapidity variable which
incorporate unitarity, analyticity and crossing \cite{K-T-T-W}.

The second surprise is that the Fourier transforms of the wedge
algebra-generating tempered PFGs are identical to operators introduced at the
end of the 70s by the Zamolodchikovs (their algebraic properties were spelled
out in more detail by Faddeev). Although the usefulness of this new algebraic
structure in the bootstrap formfactor program was immediately recognized, its
conceptual position within QFT was not clear since despite the similarity of
these objects to free field creation/annihilation operators the Z-F operators
are distinctively different from those of the incoming or outgoing free fields
of scattering theory. In the simplest case the Z-F algebra relation are of the
form ($\theta$= momentum space rapidity)%

\begin{align}
Z(\theta)Z^{\ast}(\theta^{\prime})  &  =S^{(2)}(\theta-\theta^{\prime}%
)Z^{\ast}(\theta^{\prime})Z(\theta)+\delta(\theta-\theta^{\prime}%
)\label{PFG}\\
Z(\theta)Z(\theta^{\prime})  &  =S^{(2)}(\theta^{\prime}-\theta)Z(\theta
^{\prime})Z(\theta)\nonumber\\
\phi(x)  &  =\frac{1}{\sqrt{2\pi}}\int(e^{ip(\theta)x(\chi)}Z(\theta
)+h.c.)d\theta\nonumber
\end{align}
where the notation for the structure functions $S^{(2)}$ already preempts
their physical interpretation as the two-particle S-matrix (which can be
derived \cite{Lech}). The last line defines a covariant field which, although
not being pointlike local, turns out to be wedge-like localized\footnote{The
$x$ continues to comply with the covariant transformation law but it is not a
point of localization i.e. the smearing with wedge supported test functions
$\phi(f)$ does not lead to an improvement in localization if one reduces the
support of $f.$} i.e. it commutes with its ``modular opposite'' $J\phi(x)J$
where $J=S_{scat}J_{0},$ and $S_{scat}$ is the scattering operator in Fock
space and $J_{0}$ the TCP operator of the free incoming particles$.$ This
interpretation of the Z-F algebra operators in terms of localization concepts
turns out to be a valuable guide for the construction of tighter localized
algebras $\mathcal{A}(D)$ associated with double cone regions by computing
intersections of wedge algebras whose generating operators turn out to be
infinite series in the $Z^{\prime}s$ with coefficient functions which are
generalized formfactors. The problem of demonstrating the existence of a
nontricial QFT associated with the algebraic structure (\ref{PFG}) of the
wedge algebra generators is then encoded into a nontriviality statement
($\mathcal{A}(D)\neq C\mathbf{1}$) for the double cone intersections; the fact
that the Z-F algebra is different from that of free field
creation/annihilation operators has the consequence that the operators in the
intersection have infinitly many vacuum polarization components (connected
formfactors) involving all particle numbers. \textbf{In this way the problem
of existence of nontrivial QFTs becomes disconnected from the inexorable
short-distance problems of the standard approach }\cite{Sc3}.

Factorizing models are presently the testing ground for new ideas on
the\ age-old unsolved problem of existence of interacting QFT \cite{Lech} i.e.
on whether the principles of QFT and the concepts used to implement them
continue to be mathematically consistent in the presence of interactions. It
is remarkable that this construction program leads to a derivation of those
recipes (crossing for formfactors kinematical pole relation,...) which were
used in a more or less ad hoc fashion \cite{Smir} in the standard formulations
of the bootstrap-formfactor program from first principles.

It is profitable to pause for a moment and ask the question in what sense
these findings are related to the Coleman-Mandula theorem which states (on the
basis of analytic properties of the scattering matrix) that apart from the
case of graded symmetries which needs a special consideration, a non-trivial
entanglement (trivial = tensor product) of spacetime with inner symmetry for
theories with a mass gap is only possible in d=1+1. The main theorem in
\cite{B-B-S} further qualifies this statement by linking it to the existence
of tempered PFGs which are a generalization of free fields. Free fields and
tempered PFGs have in common that they possess in addition to Poincar\'{e}
invariance infinitely many additional other conserved charges which are not
really inner. But to cite massive free field as examples for a nontrivial
entanglement of spacetime symmetry with something else is a bit awkward from a
physical viewpoint and this extends to the factorizing model. It would be
physically more appropriate to say that the two-dimensional exception allowed
by the Coleman-Mandula is related to special vacuum polarization properties
expressed by the existence of tempered wedge-localized PFGs. The vacuum
polarization properties are fundamental on the quantum level whereas the
presentation in terms of integrability in the sense of infinitely many
conserved charges is a formulation which requires classical hindsight.

The recognition that the knowledge of the position of a wedge-localized
subalgebra $\mathcal{A}(W)$ with $\mathcal{A}(W)^{\prime}=\mathcal{A}%
(W^{\prime})$ within the full Fock space algebra $B(H)$ together with the
action of the of the Poincar\'{e} group in $B(H)$ on the $\mathcal{A}(W)$)
determines the full net of algebras $\mathcal{A}(\mathcal{O})$ via
intersections%
\begin{equation}
\mathcal{A}(\mathcal{O}):=\bigcap_{W\supset\mathcal{O}}\mathcal{A}(W)
\end{equation}
is actually independent of spacetime dimensions and factorizability. But only
in d=1+1 within the setting of factorizable models one finds simple generators
for $\mathcal{A}(W)$ which permit the computation of intersections. Outside of
factorizing models, wedge generators cannot be expected to have such a simple
relation to in/out fields and one may have to take recourse to a perturbative
approach, starting with the free fields as wedge generators and taking for
$S_{scat}$ (which fixes the position of the commutant) a lowest order tree
graph expression and afterwards recursively computing corrections to the wedge
generators and $S_{scat}$ in the spirit of the Epstein-Glaser iteration using
the fact that $S_{scat}$ enters the definition of the commutant. Since one is
not aiming at a perturbation theory of pointlike localized fields but rather
of wedge-localized generators, there should be no short distance problem
(inasmuch as there is non in the Z-F wedge generators); in this way one may
have the chance to see the \textbf{true intrinsic frontiers of QFT} according
to its own physical principles (beyond those created by the use of short
distance singular pointlike localized fields) 

\section{Wedge localized algebras and holographic lightfront projection}

With the particle picture outside factorizing theories being made less useful
by de-localization through interaction induced vacuum polarization, it is
encouraging to note the existence of another constructive idea also based on
modular inclusion and intersections which does not require the very
restrictive presence of wedge-localized PFGs.. This is the holographic
projection to the lightfront. In d=1+1 it maps a massive (non-conformal) QFT
to a chiral theory on the lightfront (lightray) $x_{-}=0$ in such a way that
the global ambient algebra on Minkowski spacetime $\mathcal{A}(M)=B(H)$ and
its global holographic lightfront projection $\mathcal{A}(LF)=B(H)$
coalesce\footnote{Lightfront holography also works for higher-dimensional
conformal theories, the d=1+1 conformal models are the only exception (already
their classical version requires the knowledge of both upper and lower
boundary characteristic data to fix the wave function inside the wedge).}, but
the local substructure (the spacetime-indexed net) is radically different; the
only algebra (besides the global) which is shared between the lightfron
spacetime indexing and the ambient spacetime indexing is the wedge-localized
algebra which is identical to the algebra of their upper lightfront boundary
$\mathcal{A}(W)=\mathcal{A}(LF(W))$.

Using concepts of modular theory (modular inclusions and modular intersections
of wedge algebras) one can construct the local structure \cite{Sc3} (i.e. the
local algebraic net) and identify the subgroup $G(LF)$ of the Poincar\'{e}
group which is the symmetry group of the holographic lightfront projection.
Whereas some of the ambient Poincar\'{e} symmetries are evidently lost (in
d=1+1 the translation leading away from the lightray), the holographic
projection is also symmetry-enhancing in the sense that the rotational
symmetry of the Moebius group associated with the compactified lightray (and
according to subsection 4 also the infinite dimensional Diff(S$^{1}$) group)
becomes geometric. These symmetries are already present in the ambient theory,
but they are not noticed because they acts there in a nonclassical fuzzy
manner and hence escapes the standard quantization approach \cite{Sc3}.

Whereas the holographic lightfront projection exists in every spacetime
dimension, the setting of d=1+1 factorizing models presents a nice theoretical
laboratory to study the intricate exact relation between massive models and
their chiral projection in the context of mathematically controllable
surrounding. Those chiral observables, which appear as the holographic
projection of factorizable massive models, have the property of admitting
generators with simple Z-F algebraic creation/annihilation properties and a
covariant transformation property under the full two-dimensional Poincar\'{e}
group. It is clear that a chiral theory specified in terms of such P-covariant
operators leads (in analogy to free fields) a unique natural
\textit{holographic inversion} (but without guaranty of its existence) from a
chiral theory to a massive two-dimensional ambient theory. But not having
access to this additional knowledge, the relation of ambient theories to their
holographic projection is not expected to be one-to-one. As in the case of the
canonical equal time formalism, one rather expects that the specification of a
kind of ``Hamiltonian'' propagating\textit{ in the }$\mathit{x}_{-}$\textit{
direction} is necessary for a unique holographic inversion.

The holographic relation between chiral models and factorizing theories is
different from Zamolodchikov's perturbative identification and classification
of factorizing theories starting from a perturbation of chiral models changes.
In the latter case the representation space of the zero mass limit deviates
from that of the ones in which the different members of the original massive
universality class live. The specification of the chiral perturbation in
conjunction with the restriction to the factorizable members of the
universality class may make them singletons, but they nevertheless are
different theories living in different Hilbert spaces. Zamolodchikov's
successful approach is not based on any one-to one correspondence between
massless fields and their would be massive counterpart i.e. there is no ``mass
dressing operation''\footnote{The number of massless fields in their local
equivalence class is smaller since some massive fields vanish in the limiting
theory as can be shown in examples.}. What matters is that those fields which
are lost in the chiral limit are composites of those massive fields which
persist in that limit. In holography on the other hand there is no loss of
algebraic structure, only a radical change of spacetime indexing of the
algebraic substrate between the original ambient theory and its lightfront
holographic projection. An intuitive useful analogy is that to stem cells
which can be grown into different organs; the abstract algebraic substrate
(e.g. the abstract Weyl algebra) can be converted into different
spacetime-indexed algebraic nets. It is interesting to note that this picture
is precisely the idea which underlies the recently discovered local covariance
principle for QFT in curved spacetime \cite{B-F-V}.

It already was alluded to that the entire issue of statistics of particles
looses its physical relevance for 2-dim. massive models; they can be changed
without affecting the physical content \cite{schizon}. Instead such notions as
order/disorder fields and solitons take their place. In such a world it would
be possible to rewrite a Mendeleev periodic table of elements in terms of
Bosons. As the mass approaches zero ($\sim$ short distance limit) ``confined''
charges become liberated and the situation changes to the one in conformal
theory where the chiral (possibly anomalous) spins (= dimensions) are uniquely
related to the statistic (commutation relations) by the conformal
spin-statistics theorem.

Since the classification of Z-F algebras is a structurally simpler (possibly
computationally more complicated) program than that of chiral observable
algebras, it may very well turn out that method of holographic lightray
projection of factorizable theories may be useful for a more intrinsic
constructive approach to chiral observable algebras.

There are many additional observations on factorizing models which, although
potentially important for more profound understanding of QFT (e.g.
renormalization group flows, the meaning of the c-parameter in energy-momentum
commutation relations outside of the chiral setting, the thermodynamic Bethe
Ansatz) which have not yet reached their final conceptual placement which
identifies them as special two-dimensional manifestations of general concepts
of QFT.

\section{Concluding remarks}

In order to present two-dimensional models as a theoretical testing ground for
the still unfinished project of QFT (which was initiated more than three
quarters of a century ago by Pascual Jordan's ``Quantelung der Wellenfelder''
\cite{Schweber}, and in particular to illustrate his later plea for a
formulation without ``classical crutches''), we have used the three oldest
models proposed by Jordan, Lenz-Ising and Schwinger as paradigmatic role
models. The conceptual messages they reveal allow to analyze and structure the
vast contemporary literature on low dimensional QFT and expose the
achievements as well as the unsolved problems in a comprehensible manner
without compromising their depth and complexity (which even their protagonists
were not aware of).

The new way of viewing QFT with the help of modern developments in algebraic
QFT tested in the setting of 2-dim. QFT becomes most apparent if one looks at
the changes in the way one treats the problem of proving the existence of
models of QFT. The old measure theoretical approach ($P\Phi_{2}$) required
intermediate regularizations and was limited by the requirement of short
distance behavior to low-dimensional models. It shares with the new approach
the presence of some free field or free field-like reference structure from
which the construction starts (e.g. the tensor product reduction of free
massless Dirac fields for level k current algebras, the Zamolodchikov-Faddeev
algebra creation/annihilation operators for factorizing models0. But in
contrast to the old constructive QFT the free reference structure becomes
modified by a sequence of very nontrivial steps (tensor product reduction,
coset- and orbifold construction, extensions and intersections of operator
algebras) which finally lead to very nontrivial objects whose short distance
behaviour is far removed from that of the starting free fields, so that the
original auxiliary Fock space structure is of not much use and becomes
replaced by a reduced Hilbert space description which is more intrinsic to the
resulting algebraic structure. In this process of construction one learns more
about the physical content of models than in the old approach to constructive
QFT which was limited to near canonical dimensions. Different from this old
constructive approach \cite{Gl-Ja}, there is never any need to go outside the
principles of local QFT in intermediate steps (as regularization of short
distance singularities, recovering Poincar\'{e} invariance only in the
infrared limit on functional measures); to the contrary, the modular-based
construction depends entirely on maintaining sharp covariant localization
properties in every step of the computation. As a result the old plagues of
short distance divergencies and their control are gone and instead one has to
face the new problem of deciding whether certain intersections of operator
algebras are nontrivial ($\neq C\mathbf{1}$). Even the use of singular
pointlike fields in the bootstrap-formfactor program does not cause any short
distance problem as long as one only works with formfactors and avoids
correlation functions. An intrinsic indication of the distance to free
theories (i.e. the presence of interactions) is the interaction-caused vacuum
polarization\footnote{The fact that the S-matrix of these models lacks real
particle creation does not improve their short distance field correlation
since the latter are determined by interaction-induced vacuum polarization
(``virtual'' particle creation).} which entails the absence of
subwedge-localized polarization-free one-particle states in the setting of
massive factorizing models. The very existence of models whose local algebras
only admit generating fields whose short distance singularities are worse than
those of the superrenormalizable $P\Phi_{2}$ models shows that the ultraviolet
problems of QFT are not intrinsic but were forced upon quantum field theorists
because they entered QFT via (Lagrangian) quantization and had to do their
calculations with rather singular pointlike covariant fields and their
correlation functions\footnote{String theorists who criticise QFT on the basis
of ultraviolet divergencies in reality only criticize the specific
computational use of singular pointlike fields and their correlations. Their
criticism does not go to the intrinsic content.}. The new approach to QFT
should aim at a ``modular perturbation theory`` for generators of
wedge-localized algebras and come to tighter localizations by the process of
intersections, with the pointlike generators being a convenient
coordinatization for the computed net but not to be used via their correlation
functions in the actual calculations. Such an approach should reproduce the
results of standard approach in case of renormalizable interactions (in the
standard sense), and, what is more important, by problematizing old recipes it
should tell us something about the true frontiers of perturbative iterations
which are set by the principles of localizable QFT. In this context it is
interesting to observe that already the encoding of (m,s) Wigner
representation into modular semiinfinite string-localized fields (still
singular objects) already improves the short-distance behaviour by
transferring part of the singularity to the fluctuating spacelike direction (a
localization point in a one-dimension lower de Sitter spacetime) \cite{M-S-Y2}.

It is expected that two-dimensional QFTs will continue to play a crucial role
in the future development of these new aspects of QFT either directly (e.g.
holographic projections$\rightarrow$ generalized chiral theories) or more
indirectly as a testing ground for new concepts.

\end{document}